\documentclass[preprint,showpacs,preprintnumbers,amsmath,amssymb]{revtex4}

\usepackage{graphicx}
\usepackage{dcolumn}
\usepackage{bm}

\begin{document}

\title{The onset of superfluidity in capillary flow of liquid helium 4}

\author{Shun-ichiro Koh}
 
 \email{koh@kochi-u.ac.jp}
\affiliation{ Physics Division, Faculty of Education, Kochi University  \\
        Akebono-cho, 2-5-1, Kochi, 780, Japan 
}%

\date{\today}

\begin{abstract}
 The onset mechanism of superfluidity is examined by taking the case of 
 the capillary flow of liquid helium 4.  In the capillary flow, a substantial fall of the shear 
 viscosity $\eta $ has been observed in the normal 
 phase ($T_{\lambda}<T<3.7K$). In this temperature region, under 
 the strong influence of Bose statistics, the coherent many-body wave 
 function grows to an intermediate size between a macroscopic and a 
 microscopic one, which is different from thermal fluctuation.  
 We consider such a capillary flow by including it in a general picture 
 that includes the flow of rotating helium 4 as well. Using the  Kramers-Kronig relation,
we express $1/\eta $ in terms of the generalized susceptibility of the 
system, and obtain a formula for the shear viscosity in the vicinity 
 of $T_{\lambda}$.  Regarding  bosons without the condensate as a 
 non-perturbative state, we make a perturbation   
 calculation  of the susceptibility with respect to the repulsive 
 interaction. With decreasing temperature from $3.7K$, the 
 growth of the coherent wave function gradually suppresses the shear 
 viscosity, and makes the superfluid flow stable. Comparing formulas 
 obtained to the experimental data, we estimate that the ratio of the superfluid density $\hat 
 {\rho¥}_s(T)/\rho$  defined in the mechanical sense reaches $10^{-5}$ just above $T_{\lambda}$.

 \end{abstract}

\maketitle

\section{\label{sec:level1}Introduction}
  Superfluidity was first discovered as a frictionless flow through a 
  capillary  or a narrow slit  \cite {kap}.  
The simplest experiment for this phenomenon is as follows. A
capillary is set up vertically in gravity, the upper 
end of which is connected to a reservoir, and the lower end of which is 
open to a helium bath \cite {men}\cite {tje}\cite {zin}. 
After some liquid is filled into the reservoir, the liquid flows out of 
the reservoir through the capillary. In gravity, the level $h(t)$ of the 
liquid in the reservoir varies with time $t$ as $h(t)=h(0)e^{-\alpha t}$, where 
$\alpha$ is a constant that is inversely proportional to the coefficient of 
shear viscosity $\eta$.  Figure 1 shows the kinematic shear viscosity 
$\nu (T)=\eta (T)/\rho (T)$ of liquid helium 4 in the vicinity of 
$T_{\lambda}$  in $1$ atm ($\rho$ is a  density) \cite {bar}.  When liquid helium 4
approaches the superfluid phase ($\eta \rightarrow 0$, hence $\alpha \rightarrow 
\infty $),  it extremely rapidly flows down through the capillary.  
This phenomenon instilled us with the notion that the frictionless flow is a 
prototype of superfluidity.

We normally interpret the phenomena in liquid helium 4 using the two-fluid model, 
the essence of which is to separate the normal-fluid and superfluid  
part from the beginning, and to assume  the abrupt emergence of the superfluid part 
at the $\lambda$-point. All anomalous properties above $T_{\lambda}$ are 
regarded as being caused by thermal fluctuations.  In cooling the Bose 
systems, we observe the increase of the specific heat, or the damping 
of the sound wave at  $|T/T_{\lambda}-1|<10^{-2}$. These anomalous  
properties above $T_{\lambda}$, in which all   
particles randomly move with no specific direction, are attributed to the thermal 
fluctuation that is characteristic to the Bose systems  \cite {pha}. The wave functions 
arising from thermal fluctuation are randomly oriented.

In the capillary flow, we know a remarkable experimental result in 
connection with this point. In liquid helium 4  just above 
the $\lambda$ point, the kinematic shear viscosity $\nu (T)$ does not 
abruptly drops to zero at  $T_{\lambda}$ as in Fig.1. Rather, after 
reaching a maximum value at 3.7K, it gradually decreases with decreasing  
temperature, and finally drops to zero at the $\lambda$-point.
 This behavior has been often attributed to thermal fluctuations, but 
 this explanation is questionable.  Superfluidity, 
observed in the capillary flow or the  rotation in a bucket, requires a 
long-lived translational motion of particles with a stable specific 
direction. The characteristic time of the flow experiment is far longer 
than the relaxation time $\tau _{th}$ of 
thermal fluctuations \cite {cri}. Even if the coherent wave function obeying Bose 
statistics arises as a thermal fluctuation, it will decay long 
before it manifests itself as a macroscopic flow. Furthermore, the wave function arising from 
thermal fluctuation is randomly oriented, but the macroscopic flow must 
have a specific direction. In this respect, it is impossible that thermal fluctuations lead to 
superfluidity \cite {even}. Similarly, the macroscopic flow showing a 
partial disappearance of $\nu (T)$ at $T_{\lambda}<T<3.7K$ needs the stable   
translational motion of atoms with a specific direction, and is
qualitatively different from thermal fluctuations.

\begin{figure}
\includegraphics [scale=0.6]{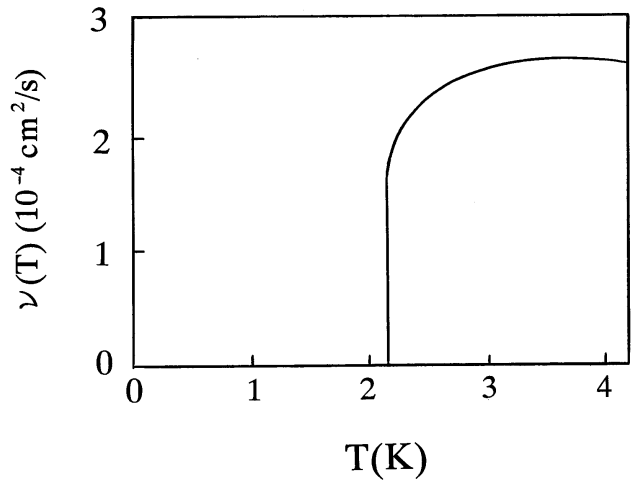}
\caption{\label{fig:epsart} 
   The temperature dependence of the kinematic shear viscosity $\nu (T)$ of liquid 
  helium 4 obtained in Ref.\cite {bar}. }
\end{figure}

 London stressed that the total disappearance 
 of shear viscosity is attributed to  $rot \mbox{\boldmath $v$}=0$ over 
 the whole volume of a liquid \cite {lod}. The partial disappearance 
 of shear viscosity at $T_{\lambda}<T<3.7K$ in Fig.1  suggests the 
 emergence of regions in which $rot  \mbox{\boldmath $v$}=0$ is locally realized. 
Compared to an ordinary liquid, liquid helium 4 above $T_{\lambda}$ 
 has a  $10^{-3}$ times smaller coefficient of shear viscosity.  
 Although in the normal phase, it is already an anomalous  
 liquid  under the strong influence of Bose statistics. 
 Hence, it seems  natural to assume the existence of large but not yet macroscopic
 coherent wave functions \cite {koh}. 
 If there exist such intermediate-sized wave functions above 
 $T_{\lambda}$, they must affect the macroscopic flow.

In conventional theories, we normally do not think of the large but not yet 
macroscopic wave function. This is because we use the infinite volume limit ($V\rightarrow 
\infty $) in order to make a clear definition of the order parameter in the phase transition.
 In the  $V\rightarrow \infty $ limit, ``large but not yet 
macroscopic'' is equivalent to ``microscopic'', and  therefore the 
 intermediate-sized wave function disappears from the beginning. 
For the system in which the distinction between ``microscopic'' and 
``macroscopic'' is clear, the $V\rightarrow \infty $ limit is a realistic 
one.  For the Bose system at low temperature, however, the Bose 
statistical coherence develops to a macroscopic or an intermediate size, and the container 
and the coherent region of a liquid are comparable in size. 
 The $V\rightarrow \infty $ limit is therefore a questionable assumption 
 for considering the mechanical properties of the Bose system just above $T_{\lambda}$. (For the 
 thermodynamical properties, we encounter no difficulty in the  $V\rightarrow \infty $ limit.)

The shear viscosity of liquid helium 4 has been subjected to considerable 
experimental and theoretical studies. These studies are mainly focused on 
the famous paradox at $T<T_{\lambda}$ that superfluid helium 4 behaves as 
a viscous or non-viscous fluid, depending on the experimental methods. 
For $\nu (T)$ at $ T_{\lambda}<T<3.7K$, it is often said that it 
 resembles $\nu (T)$ of a gas rather than a liquid in its magnitude and  
in its temperature dependence. But this explanation is somewhat misleading. 
In liquid helium 4, many features associated to Bose statistics have been masked by the 
strongly interacting nature of the liquid.  Well below the lambda  point $T_{\lambda}$, various 
excitations of liquid helium 4 are strictly suppressed except for phonons and 
rotons.  Hence, although the excitation in a liquid, these 
phonons and rotons are normally regarded as a weakly interacting dilute Bose gas \cite {kha}.  
 For $\nu (T)$ above $T_{\lambda}$, however, the 
dilute-gas picture has no grounds, because the macroscopic 
condensate, which is the basis for the above picture, has not yet developed. 
 To explain $\nu (T)$ above $T_{\lambda}$, we must deal with the 
 dissipation mechanism of a liquid and the influence of Bose statistics on it.

 At present, for the accuracy of data, the early measurements of $\eta (T)$ using the  
capillary have been superseded  by new  measurements using more accurate 
techniques such as the vibrating wire \cite  {goo}\cite {wel} \cite {bru}.
 Above the $\lambda$ point, the latter method gives not only   
 quantitatively similar, but more precise data of $\eta (T)$. Applying the 
 statistical analysis to these accumulated data,  
 a precise temperature dependence of $\eta (T)$ and $\nu (T)$ above $T_{\lambda}$ was 
obtained \cite {bar}. A theory worth comparing with these precise data is desired.

This paper gives a model of the  onset of superfluid flow in the capillary.   
Instead of assuming the macroscopic condensate from the beginning,  
we begin with the repulsive Bose system with no off-diagonal-long-range order (ODLO), and  
show that, with decreasing temperature  above $T_{\lambda}$, the coherent many-body wave 
function  gradually grows. In Ref.\cite {koh}, we showed that this growth reflects in the rotational 
properties such as the moment of inertia above $T_{\lambda}$.
 In contrast with the rotation, however, the capillary flow of a classical liquid is a 
 phenomenon that is accompanied by thermal dissipation.   
The frictionless capillary flow  at $T<T_{\lambda}$ is superfluidity 
which occurs in the strongly dissipative system when it is at 
$T>T_{\lambda}$. Hence, to consider this phenomenon, we must devise a different formalism. 
To take the nature of the liquid into consideration, we will go back to 
fluid mechanics in the normal  phase, and begin with  
the solution of the Stokes equation for the capillary flow (Poiseuille  
formula). The capillary flow is embedded into a general  
picture that is applicable both to dissipative and to non-dissipative 
flows using the generalized susceptibility.  
Applying the Kramers-Kronig relation to the generalized susceptibility, we relate the 
capillary flow to its non-dissipative counterpart, rotating helium 4 in a 
bucket, and derive  the decrease of $\nu (T)$ from the growth of the coherent 
wave function. When this result is combined with that in Ref.\cite {koh},
 we come to understand the onset of two types of superfluidity in the 
 shear viscosity and in the moment of inertia from a common origin.

This paper is organized as follows.  Section.2 examines the capillary 
flow in terms of the generalized susceptibility.  
 Using this method, Sec.3A gives a formula for the shear viscosity in the 
 vicinity of $T_{\lambda}$, and Sec.3B examines the role of Bose statistics and the 
repulsive interaction in the suppression of the shear viscosity. Section.4 develops a 
microscopic model of the onset of the superfluid flow.  By generalizing the formalism in 
 Ref.\cite {koh} to the time-dependent case, we  derive the 
decrease of $\eta $ just above $T_{\lambda}$ from the growth of the 
coherent wave function, and discuss the stability of superfluid flow through a capillary. 
To describe the system just above $T_{\lambda}$, the mechanical superfluid density $\hat 
{\rho¥}_s(T)$, which is defined without using $V\rightarrow \infty$, is 
more useful than the  conventional superfluid density $\rho¥ _s(T)$. 
In Sec.5, using $\nu (T)$ of Fig.1, we estimate the ratio $\hat {\rho _s}(T)/\rho$,
and compare this $\hat {\rho _s}(T)/\rho$ with another $\hat {\rho 
_s}(T)/\rho$ obtained by the rotation experiment \cite {hes}.  
 Section.6 discusses some related problems.

\section{Capillary flow}

 In an ordinary liquid flowing along x-direction, 
the shear viscosity causes the shear stress $F_{xy}$ between two adjacent 
layers at different velocities (see Fig.2)
\begin{equation}
 F_{xy}=\eta \frac{\partial v_x}{\partial y¥}¥.
	\label{¥}
\end{equation}¥
In the linear-response theory, $\eta $ of a stationary flow is given by the 
following two-time correlation function of a tensor $J_{xy}(t)=-\sum_{i}(p_{i,x}p_{i,y}/m¥)$
\begin{equation}
   \eta =\frac{1}{Vk_BT¥}¥\int_{0}^{\infty¥}dt<J_{xy}(0)J_{xy}(t)>¥.
	\label{¥}
\end{equation}¥
 In principle, $\eta$ of a liquid, and therefore the effect of Bose statistics on $\eta$ is obtained by 
calculating an infinite series of the perturbation expansion of   
Eq.(2) with respect to the particle interaction. 
However, it seems to be a hopeless attempt, because the dissipation in a liquid
 is a complicated phenomenon that allows no simple approximation \cite {han}.

  \begin{figure}
\includegraphics [scale=0.5]{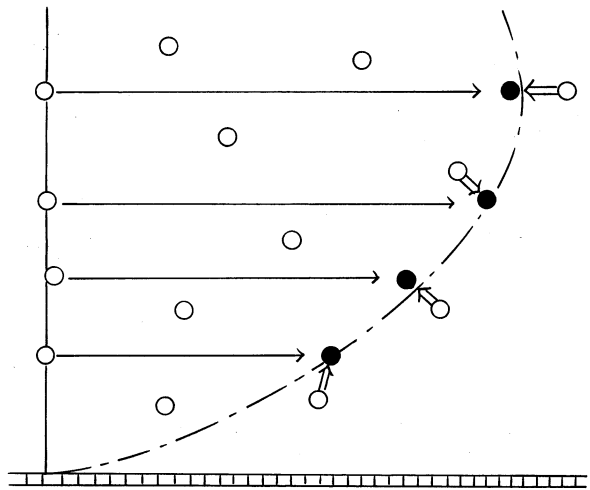}
\caption{\label{fig:epsart} A flow through a capillary.  }
\end{figure}

Fluid mechanics is a simple phenomenological theory of non-equilibrium 
behaviors for situations in which physical quantities vary slowly in space 
and time. In this paper, we will develop an intermediate theory 
between the microscopic and phenomenological one \cite {kad}, and 
derive some information on the shear viscosity from the correspondence between 
these two different levels of description. 
 In an ordinary flow through a capillary with a circular cross-section (a radius $d$ and a length 
 $L$), the Stokes equation has a solution of the velocity distribution $v_z(r)$ under the 
 pressure difference $\Delta P$ such as 
\begin{equation}
 v_z(r)=\frac{d^2-r^2}{4\eta¥}¥\frac{\Delta P}{L¥}¥,
	\label{¥}
\end{equation}¥
where $r$ is a radius in the cylindrical coordinates  (Poiseuille flow \cite {lan}. Fig.2).
 Without loss of generality, we may focus on a flow velocity on the axis 
 of rotational symmetry (z-axis).  We define  a mass-flow density 
 $\mbox{\boldmath $j$}=\rho\mbox{\boldmath $v$}(r=0)$, and rewrite Eq.(3) as
\begin{equation}
 \mbox{\boldmath $j$}=-\sigma d^2\frac{\Delta \mbox{\boldmath $P$}}{L¥},  \qquad  \sigma =\frac{\rho}{4\eta¥}¥.
	\label{¥}
\end{equation}¥
We call $\sigma$ {\it the conductivity of a viscous liquid flow through a capillary\/}
 ($\mbox{\boldmath $P$}=P\mbox{\boldmath $e$}_z$)  \cite {cur}. 
  From a microscopic view point, it is a long-standing problem 
to derive the absolute value of $\eta$ using Eq.(2) \cite{jeo}. 
In this paper, however, instead of Eq.(2), we will regard Poiseuille's formula Eq.(4) 
as a starting point.  We regard Eq.(4) as a phenomenological 
linear-response relation, and compare it to the microscopic 
formula for the mass-flow density $ \mbox{\boldmath $j$}$ like Eq.(28). 
By this comparison, we will extract only the change of $\eta$ near 
$T_{\lambda}$ from the total $\eta$ in Sec.3, and   
derive this change from a microscopic model in Sec.4.

In a flow through a capillary set up vertically, the level of a liquid in 
the reservoir varies with time $t$ as $h(t)=h(0)e^{-\alpha t}$ (see Appendix.A). 
Hence,  $\Delta P(t)$ in Eq.(4) decreases as $\Delta 
P(t)=\Delta P(0)e^{-\alpha t}$ as well.  With the aid of   
 \begin{equation}
	e^{-\alpha t}=\frac{1}{\pi¥}¥\int_{0}^{\infty¥}¥\frac{1}{\sqrt{\omega ^2+\alpha ^2¥}¥}¥e^{i(\omega t-\phi )}d\omega,
	\label{¥}
\end{equation}¥
($\tan\phi =\omega/\alpha$), $\Delta \mbox{\boldmath $P$}(t)$ is 
decomposed into the frequency components $\Delta \mbox{\boldmath 
$P$}(\omega)$ within a range of $[0,\omega _e]$ where $\omega _e$ is a frequency width. 
 Let us  generalize Eq.(4) to the case of $\mbox{\boldmath $j$}(\omega)e^{i\omega t}$ under 
 the oscillatory pressure $\mbox{\boldmath $P$}(\omega)e^{i\omega t}$ as follows  
\begin{equation}
 \mbox{\boldmath $j$}(\omega)=-\sigma (\omega) d^2\frac{\Delta \mbox{\boldmath $P$}(\omega)}{L¥}.
	\label{¥}
\end{equation}¥
The conductivity spectrum $\sigma (\omega)$  must 
 satisfy the following sum rule \cite {sum}
\begin{equation}
   \frac{1}{\pi ¥}¥\int_{0}^{\infty¥}\sigma (\omega)d\omega¥=f(d)¥,
	\label{¥}
\end{equation}¥
where  $f(d)$ is a conserved quantity. (The Stokes equation gives an 
expression of $\sigma (\omega)$ and $f(d)$ of the classical liquid. See Appendix.B.) 
In Eq.(4), we must incorporate the effect of time variation of
 $\Delta \mbox{\boldmath $P$}(t)$ into $\mbox{\boldmath $j$}$ and 
 $\sigma$, and we define the frequency-averaged $\sigma$  
 as $\sigma=\omega _e^{-1}\int_{0}^{\omega _e¥}¥\sigma (\omega)d\omega$ 
using $\sigma (\omega)$ in Eq.(6) .

 The change of the starting point from Eq.(2)  to Eq.(4) permits us to relate 
 different types of superfluidity: (a) the frictionless capillary flow, and (b) the 
 nonclassical flow in a rotating bucket.  The phenomenological 
 equation (6) is appropriate to be generalized so as to include different 
types of superfluidity.  Such a method originated in studies of electron 
superconductivity \cite {fer}. Between the electrical conduction and the 
Meissner effect in superconductivity,  we find a relationship parallel to 
the relationship between the capillary flow and the rotational flow in superfluidity,

The generalization of Eq.(6) is as follows. 
The conductivity spectrum $\sigma (\omega)$, which describes the 
dissipative response of a liquid, corresponds to an imaginary part of the 
generalized susceptibility $\chi (\omega)$ that describes both dissipative 
and non-dissipative responses of a liquid.  We will generalize Eq.(6) so 
as to include not only a response in phase with $\Delta \mbox{\boldmath $P$}(\omega)$ 
but a response  out of phase. Hence, $\sigma (\omega)$ in Eq.(6) becomes 
 a complex number $\sigma_1+i\sigma_2$ as follows
\begin{equation}
 \mbox{\boldmath $j$}(\omega)=-\left[\sigma_1(\omega)+i\sigma_2(\omega)\right]¥d^2 \frac{\Delta \mbox{\boldmath $P$}(\omega)}{L¥},
	\label{¥}
\end{equation}¥
where $\sigma (\omega)$ in Eq.(6) is replaced by $\sigma_1(\omega)$.  The 
application of the pressure gradient can be fitted into the general 
picture as follows.  As an external field, we assume a fictitious 
velocity $\mbox{\boldmath $v$}(t)$ that satisfies the  
 equation of motion  \cite {vec} 
\begin{equation}
      \rho\frac{d \mbox{\boldmath $v$}(t)}{d t¥}¥= -\frac{\Delta \mbox{\boldmath $P$}(t)}{L¥} ,
	\label{¥}
\end{equation}¥
and replace $\Delta \mbox{\boldmath $P$}(\omega)/L$ in Eq.(8) with  
 $-i\rho\omega\mbox{\boldmath $v$}(\omega)$ using Eq.(9).  
As a result, the real and imaginary part of $\sigma (\omega)$ are interchanged \cite {vel}, 
\begin{equation}
 \mbox{\boldmath $j$}(\omega)=\rho\left[-\omega\sigma_2(\omega)+i\omega\sigma_1(\omega)\right]¥¥d^2
                                                \mbox{\boldmath $v$}(\omega).
           \label{¥}
\end{equation}¥
Equation (10) is our desired formula, the coefficient of which  
  consists not only of an imaginary part $i\omega\sigma_1(\omega)$ for 
  the dissipative flow, but of a real part $-\omega\sigma_2(\omega)$ 
  for the non-dissipative one, thus forming the generalized 
  susceptibility.  According to the choice of  
  $\mbox{\boldmath $v$}(\omega)$, Eq.(10) expresses different types of
 flow, but the generalized susceptibility  
 has a general structure independent of the flow type. This means that by 
 relating $\sigma_1(\omega)$ to $\sigma_2(\omega)$, we can derive 
 $\sigma _1(\omega)$ of a capillary flow from $\sigma _2(\omega)$ of a non-dissipative flow. 
Causality requires that fluid particles begins to flow only after the pressure is applied. 
We can use the following Kramers-Kronig relation for $-\omega\sigma_2(\omega)$ and 
$\omega\sigma_1(\omega)$ in Eq.(10) 
\begin{equation}
   \sigma_1(\omega')=\frac{2}{\pi}¥\int_{0}^{\infty¥}d\omega\frac{\omega\sigma_2(\omega)}{\omega^2-\omega'^2¥}¥,
	\label{¥}
\end{equation}¥
\begin{equation}
    \sigma_2(\omega')=-\frac{2\omega'}{\pi}¥\int_{0}^{\infty¥}d\omega\frac{\sigma_1(\omega)}{\omega^2-\omega'^2¥}¥.
	\label{¥}
\end{equation}¥
If one determines $\sigma_2(\omega)$ in the non-dissipative flow, 
 one can obtain $\sigma_1(\omega)$  using Eq.(11), hence  $\sigma (\omega)$ in Eq.(6).

  As a non-dissipative flow, we will  consider the flow in a rotating bucket.
In a rotating bucket, a liquid makes the rigid-body 
 rotation  owing to its viscosity.  The rotational velocity, which is 
 used as $\mbox{\boldmath $v$}(\omega)$ in Eq.(10), is a product 
 of the angular velocity $\Omega$ and the radius $\mbox{\boldmath $r$}$  such as  
  $\mbox{\boldmath $v$}_d( \mbox{\boldmath $r$})\equiv \mbox{\boldmath 
 $\Omega$}\times \mbox{\boldmath $r$}$. For this $\mbox{\boldmath 
 $v$}_d( \mbox{\boldmath $r$})$, the dissipation function \cite {lan} 
\begin{equation}
	\Phi (\mbox{\boldmath $r$})=2\eta \left(e_{ij}-\frac{1}{3¥}e_{kk}\delta_{ij}\right)^2¥
	\label{¥}
\end{equation}¥
is zero at every $r$, where $2e_{ij}=\partial v_i/\partial x_j+\partial v_j/\partial x_i$. 
 This means that, except at the boundary  to the wall, there is no  
 frictional force within a liquid even in the normal phase, and the rigid-body 
 rotation is therefore a non-dissipative flow.  
 (On the other hand, for the Poiseuille flow Eq.(3), Eq.(13) is not zero 
 at every $r$ except for $r=0$.  Fluid particles in the capillary flow experience thermal 
dissipation not only at the boundary, but within the flow.)

The flow in a rotating bucket is formulated using the generalized susceptibility of 
the system  $\chi(\mbox{\boldmath $r$},\omega)$ \cite {noz} (see Appendix.C), 
which is decomposed to the longitudinal and transverse part ($\mu , \nu =x,y,z$)
\begin{equation}
	\chi_{\mu\nu}(q,\omega )=\frac{q_{\mu}q_{\nu}}{q^2¥}\chi^L(q,\omega)     
	            +\left(\delta_{\mu\nu}-\frac{q_{\mu}q_{\nu}}{q^2¥}\right)¥\chi^T(q,\omega) .
	\label{¥}
\end{equation}¥
In the flow in a rotating bucket, the influence of the bucket propagates 
along the radial direction from the wall to the center, which is perpendicular 
to the particle motion driven by rotation (Fig.2 of Ref.\cite {koh}). Hence, the  
flow in a rotating bucket is a transverse response of the system described by the transverse 
susceptibility $\chi^T(q,\omega)$. In the right-hand side of Eq.(10), the 
real part of the susceptibility $-\rho\omega\sigma_2(\omega)d^2$ is expressed as
\begin{equation}
   -\rho d^2\omega\sigma_2(\omega)=\lim _{q\rightarrow 0}\chi^T(q,\omega) ¥.
	\label{¥}
\end{equation}¥

Now, we can express the conductivity of the capillary flow in terms of the 
susceptibility of the system. Using Eq.(15) in the right-hand side of Eq.(11),
 we obtain $\sigma_{1}(\omega)$ for the capillary flow 
\begin{equation}
   \rho d^2\sigma_{1}(\omega')= -\frac{2}{\pi}¥¥
         \int_{0}^{\infty¥}d\omega\frac{\lim _{q\rightarrow 0}\chi^T(q,\omega)}{\omega^2-\omega'^2¥}¥.
	\label{¥}
\end{equation}¥
 Quantum mechanics states that, in the decay from an excited state with an 
 energy level $E$ to a ground state with $E_0$, the higher excitation 
 energy $E$ causes the shorter relaxation time $\tau $, $\hbar /\tau ¥ \simeq |E-E_0|$.
  In Eq.(16),  the left-hand side includes the relaxation time $\tau $ 
 in $\sigma _{1} =\rho /(4\eta)=\rho /(4G\tau)$, whereas  
 the right-hand side  includes the excitation spectrum in 
 $\chi^T(q,\omega)$. (We used Maxwell's relation $\eta =G\tau$ \cite {max}. See Appendix.D.)  
 In this sense,  Eq.(16) is a many-body theoretical expression of $\hbar /\tau ¥ \simeq |E-E_0|$.

\section{Shear viscosity of a superfluid}

\subsection{Phenomenological argument}

In the superfluid phase, even when the pressure difference vanishes 
($\Delta P=0$) in Eq.(4), one observes a stable flow, hence $\eta =0$.
 The essence of superfluidity is that the normal-fluid and  superfluid 
part flows without any transfer of momentum from one to the other. Characteristic 
to the  frictionless flow in Eq.(6) is that in addition to  the normal fluid 
part $\sigma _n(\omega)$, the conductivity spectrum $\sigma (\omega)$ has 
a sharp peak at $\omega =0$
\begin{equation}
 \mbox{\boldmath $j$}(\omega)=-\left[\sigma _n(\omega)+A\delta(\omega)\right]¥¥d^2 \frac{\Delta \mbox{\boldmath $P$}(\omega)}{L¥},
	\label{¥}
\end{equation}¥
where $A\delta(\omega)$ is a simplified expression of the sharp peak at 
$\omega =0$ with an area $A$ \cite {fer}.  Figure 3 schematically illustrates such a change of 
$\sigma (\omega)$ when the system passes $T_{\lambda}$. The separation of 
the sharp peak from $\sigma _n(\omega)$ guarantees the absence of the 
momentum  transfer (except for $\sigma _n(\omega)$ near $\omega =0$ due to the acoustic phonon). 
When $\sigma (\omega)$ has a form of  $\sigma _n(\omega)+A\delta(\omega)$, 
it affects the conductivity $\sigma=\omega _e^{-1}\int_{0}^{\omega 
_e¥}¥\sigma (\omega)d\omega$, and $\eta $ in Eq.(4) is given by  
\begin{equation}
   \eta (T)=\frac{1}{4¥}¥\frac{\rho}{\displaystyle{\omega _e^{-1}\left(\int_{0}^{\omega 
                                    _e¥}¥\sigma _n (\omega)d\omega +A\right)¥}¥}¥¥.
	\label{¥}
\end{equation}¥

\begin{figure}
\includegraphics [scale=0.5]{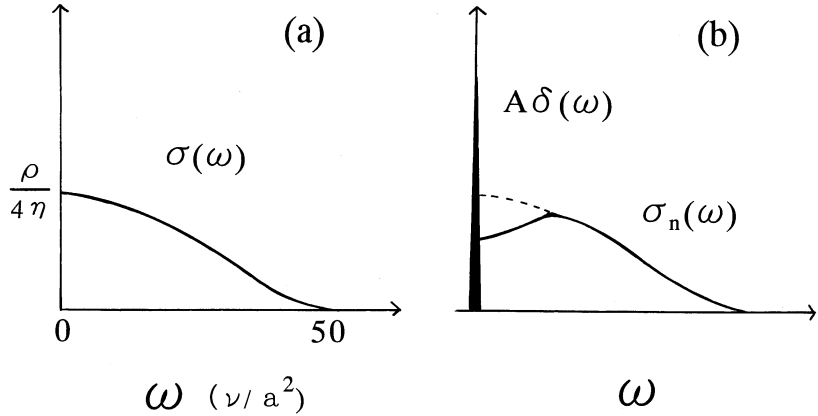}
\caption{\label{fig:epsart} The change of the conductivity spectrum 
$\sigma (\omega )$ from (a) in a classical fluid to (b) in a superfluid.  
 $\sigma (\omega )$ is given by Eq.(B6) in Appendix.B. 
   }
\end{figure}

Equation (18) is explained by the following argument on the susceptibility. 

(1) In the classical fluid, $\chi^L(q,\omega)=\chi^T(q,\omega)$ is satisfied at small $q$,  
and one can replace $\chi^T(q,\omega)$ in Eq.(16) by $\chi^L(q,\omega)$
 for a small $q$ \cite {def}.  Hence, in the classical fluid 
 the conductivity $\sigma_{1n}(\omega)$ of the capillary flow is given by 
\begin{equation}
   \rho d^2\sigma_{1n}(\omega')= -\frac{2}{\pi}¥¥
         \int_{0}^{\infty¥}d\omega\frac{\lim _{q\rightarrow 0}\chi^L(q,\omega)}{\omega^2-\omega'^2¥}¥ .
	\label{¥}
\end{equation}¥

In the superfluid phase, however, under the strong influence of Bose statistics, 
the condition of $\chi^L(q,\omega)=\chi^T(q,\omega)$  at $q\rightarrow 0$ 
is violated (see Sec.3B and 4). Hence, one cannot replace  $\chi^T(q,\omega)$ by 
$\chi^L(q,\omega)$ in Eq.(16), and must express Eq.(16) using Eq.(19) as follows 
\begin{equation}
   \sigma_1(\omega')=\sigma_{1n}(\omega')+
            \frac{2}{\rho\pi d^2¥}¥\int_{0}^{\infty¥}d\omega
            \frac{\lim _{q\rightarrow 0}[\chi^L(q,\omega)-\chi^T(q,\omega)¥]}{\omega^2-\omega'^2¥} ¥,
	\label{¥}
\end{equation}¥
in which $\lim _{q\rightarrow 0}[\chi^L(q,\omega)-\chi^T(q,\omega)¥]$ corresponds to a superfluid component.
For the later use, we define a  term proportional to $q_{\mu}q_{\nu}$ in 
$\chi_{\mu\nu}$ by $\hat{\chi}_{\mu\nu}$ as a quantity representing the 
balance between the longitudinal and transverse response as follows
 \begin{eqnarray}
       	\chi_{\mu\nu}(q,\omega)&=&\delta_{\mu\nu}\chi^T(q,\omega)
	                  +q_{\mu}q_{\nu}\left(\frac{\chi^L(q,\omega)-\chi^T(q,\omega)}{q^2¥}\right)¥ \nonumber \\ 
	                          &\equiv& \delta_{\mu\nu}\chi^T(q,\omega)+\hat{\chi}_{\mu\nu}(q,\omega). 
	\label{¥}
\end{eqnarray}¥

(2) The stability of a superfluid is measured by the value of $\chi^L(q,\omega)-\chi^T(q,\omega)$  
at the finite $\omega$. In general, the superfluid flow is not 
perfectly stable with respect to oscillating external perturbations. Above a certain 
frequency,  $\chi^L(q,\omega)-\chi^T(q,\omega)$ vanishes, and the system behaves as a classical fluid. 
(In Sec.4, we will show this using a concrete model.)

(3)  The shear viscosity near $T_{\lambda}$ is mainly determined by 
the value of $\lim _{q\rightarrow 0}[\chi^L(q,\omega)-\chi^T(q,\omega)]$ in the vicinity of $\omega =0$. 
In the right-hand side of Eq.(20), the finiteness of 
$\chi^L(q,\omega)-\chi^T(q,\omega)$ at $\omega \simeq 0$ leads to a sharp peak at $\omega '=0$, because of
\begin{equation}
  \int_{0}^{\infty¥}¥\frac{d\omega}{\omega^2-\omega'^2¥}¥=\delta(\omega ') .
	\label{¥}
\end{equation}¥
Hence, in the superfluid phase, $\sigma_1(\omega')$ shows a peculiar 
peak at $\omega' =0$ like that in Fig.3.(b).
(Since this change obeys the sum rule Eq.(7), the area $A$ of the sharp peak is equal to the area
 enclosed by a dotted and a solid curve in Fig.3(b).) 
 
(4) From now, we approximate $\chi^L(q,\omega)-\chi^T(q,\omega)$ by its value at 
 $\omega =0$ in Eq.(20), and $\sigma_1(\omega)$ has a form such as 
\begin{equation}
   \sigma_1(\omega)=\sigma_{1n}(\omega)+\frac{2}{\rho\pi d^2¥}¥
                          \lim _{q\rightarrow 0}[\chi^L(q,0)-\chi^T(q,0)]\delta(\omega). ¥
	\label{¥}
\end{equation}¥
 As the temperature decreases from $T_{\lambda}$, the area of the sharp peak increases. 
Using Eq.(23) in Eq.(18), one obtains the coefficient of shear viscosity 
\begin{equation}
  \eta (T)=\left(\frac{\rho}{4¥}\right)¥\frac{1}{\sigma_{1n}
              +\displaystyle{\frac{2}{\rho\pi d^2\omega _e¥}  
                       \lim _{q\rightarrow 0}[\chi^L(q,0)-\chi^T(q,0)]¥¥}}¥. 
	\label{¥}
\end{equation}¥
Here, we define {\it the mechanical superfluid density \/} 
 $\hat {\rho _s}(T)\equiv \lim _{q\rightarrow 0}[\chi^L(q,0)-\chi^T(q,0)]$
  ($= \lim _{q\rightarrow 0}[\hat {\chi}_{\mu\nu}(q,0)q^2/(q_{\mu}q_{\nu})]$), 
  which does not always agree with the conventional thermodynamical superfluid density 
$\rho _s(T)$.  (By ``thermodynamical'', we imply the quantity that 
remains finite in the $V\rightarrow \infty $ limit.) 

 Since the total density  $\rho $ in Eq.(24) slightly  
increases with decreasing temperature from $4.2K$ to $T_{\lambda}$ \cite {ker},  
  {\it the kinematic shear viscosity\/} $\nu (T)=\eta (T)/\rho (T)$ is 
  more appropriate than $\eta (T)$ to describe the change of the system around $T_{\lambda}$. 
We obtain a formula of $\nu(T)=\eta(T)/\rho(T)$  as follows
\begin{equation}
  \nu (T)=¥\frac{\nu _n}{1
              +\displaystyle{\frac{8}{\pi d^2\omega _e ¥} \frac{\hat 
              {\rho _s}(T)}{\rho}}\nu _n¥¥}¥, 
	\label{¥}
\end{equation}¥
where $\nu _n$ is the kinematic shear viscosity of a classical fluid, and 
satisfies $\sigma_{1n}=\rho/(4\eta _n)=1/(4\nu _n)$ in Eq.(24).   
In general, $\nu _n(T)$ of the classical liquid has the property of 
 increasing monotonically with decreasing temperature \cite {she}.
In Fig.1, $\nu (T)$ above $3.7K$ seems to show this property, but its gradual 
fall below $3.7K$ cannot be explained by the classical liquid picture,  
thereby suggesting $\hat {\rho _s}(T)\ne 0$ in Eq.(25) at $T_{\lambda}\leq T\leq 3.7K$.

In the free Bose system, using $j_{\mu}(q,\tau)$ in Eq.(C2), 
$\hat{\chi}_{\mu\nu}(q,\omega)$ in Eq.(21) has a form of 
\begin{equation}
\hat{\chi}^{(1)}_{\mu\nu}(q,\omega)  
	            =-\frac{q_{\mu}q_{\nu}}{4¥}¥
	                          \frac{1}{V¥}\sum_{p}\frac{f(\epsilon (p))-f(\epsilon (p+q))}
	                                       {\omega+\epsilon (p)-\epsilon (p+q)¥}¥,
	\label{¥}
\end{equation}¥
where $f(\epsilon (p))=\exp(\beta[\epsilon (q)+\Sigma-\mu])-1)^{-1}$ is 
the Bose distribution, $\Sigma$ is a self energy of a boson (we ignore $\omega$ 
and $p$ dependence of $\Sigma$ by assuming it small), and $\mu$ is a chemical potential.
At $T<T_{\lambda}$, $f(\epsilon (p))$ is a macroscopic 
number for $p=0$, and nearly zero for $p\ne 0$. Hence, in the sum over 
$p$, only two terms corresponding to $p=0$ and  
$p=-q$ remain. This means that $\hat{\chi}^{(1)}_{\mu\nu}(q,0)$ has a form of
 $\hat {\rho _s}(T)¥q_{\mu}q_{\nu}/q^2¥$. Practically, a nonzero $\hat 
 {\rho _s}(T)$ and a small $\omega _e$   vanishes $\nu (T)$ at 
 $T<T_{\lambda}$ in Eq.(25). When bosons form no condensate, however, the sum over $p$ in 
Eq.(26) is carried out by replacing it with an integral, and  
 $q^{-2}$ dependence of $\hat{\chi}^{(1)}_{\mu\nu}(q,0) $ disappears, 
 hence $\hat {\rho _s}(T)=0$ and $\nu (T)$ remains finite.  This means 
 that, without the interaction between particles,   
the macroscopic Bose-Einstein condensation (BEC) is the necessary condition for the decrease of $\nu (T)$.
 To explain the gradual fall of $\nu (T)$ just above $T_{\lambda}$, we must obtain  
$\chi^L(q,\omega)-\chi^T(q,\omega)$ under the particle interaction.

\subsection{The effect of Bose statistics and repulsive interaction}

One can physically explain the fall of the shear viscosity in Eq.(25) in 
terms of  Bose statistics.  When a liquid flows through a 
capillary, it moves in the same direction but with a speed that varies in 
a perpendicular direction. For the classical liquid, Maxwell obtained a simple formula 
$\eta=G\tau _{st}$ (Maxwell's relation. See Appendix.D) \cite{max}, where $G$ is the 
modulus of rigidity, and  $\tau _{st}$ is a characteristic time of 
structural relaxations that occur under the sheer stress in the flow.  (The reader must not 
confuse this $\tau _{st}$ with $\tau _{th}$ of thermal fluctuations.)
 This relation is useful for the interpretation of liquid helium 4 as well.
 In the vicinity of $T_{\lambda}$ in liquid helium 4, no structural transition is  
observed in coordinate space. Hence, $G$ may be a constant at the first 
approximation, and therefore the fall of the shear viscosity is 
attributed mainly to the decrease of $\tau _{st}$.  In view of 
$\hbar /\tau _{st}¥ \simeq |E-E_0|$, the decrease of $\tau _{st}$  suggests the  
increase of the excitation energy $E$, and it is natural to attribute it 
to Bose statistics. The relationship between  
the excitation energy and  Bose statistics dates back to 
Feynman's argument on the scarcity of the low-energy excitation in liquid helium 
4 \cite{fey1}, in which he explained  how Bose 
statistics affects the many-body wave function in configuration space. 
To the shear viscosity, we will apply his explanation.

Consider a flow in Fig.2, in which white circles represent  
an initial  distribution of fluid particles.  The long thin arrows 
represent the displacement from  
white circles on a solid straight line to black circles on a 
one-point-dotted-line curve. (The influence of adjacent layers in a viscous flow 
propagates along a direction perpendicular to the particle motion. Hence, the excitation 
associated with the shear viscosity is a transverse one.) 
Let us assume that a liquid in Fig.2 is in the BEC phase, and 
the many-body wave function has permutation symmetry everywhere in a capillary.  
At first sight, these displacements by long arrows seem to be a large-scale configuration change,  
 but they are reproduced by a set of slight displacements by short 
 thick arrows from the neighboring white circles in the initial distribution. 
In general, the transverse modulation, such as the displacement by shear 
viscosity, does not change the particle density in the large scale, and 
therefore, to any given particle after displacement, it is always 
possible to find such a neighboring particle in the initial distribution.
  In Bose statistics, owing to permutation symmetry, one cannot 
distinguish between two types of particles after displacement, one  moved 
from the neighboring position by the short arrow, and the other moved 
from distant initial positions by the long arrow. {\it Even if  the 
displacement made by the long arrows is a large 
displacement in classical statistics, it is only a slight 
displacement  by the short arrows in Bose statistics \/}. 
(In contrast, the longitudinal modulation results in the large-scale 
 inhomogeneity in the particle density, and therefore it 
 is not always possible to find such neighboring initial positions.)

Let us imagine this situation in 3N-dimensional configuration space. The 
excited state made of slight  displacements, which is characteristic of Bose 
statistics, lies in a small distance from the ground state in configuration space.  
On the other hand,  the wave function of the excited state is orthogonal to that of the 
ground state in the integral over configurations. Since 
 the amplitude of the ground-state wave function is 
 uniform in principle, the amplitude of the excited-state wave function must 
spatially oscillate around zero. Accordingly, the wave function of the 
excited state must oscillate  around zero within a small 
distance in configuration space.  The kinetic energy of the system 
is determined by the 3N-dimensional gradient of the many-body wave 
function in configuration space, and therefore  this steep rise and fall 
of the amplitude raises the excitation energy of this wave function. 
The relaxation from such an excited state is a rapid process   
with a small $\tau _{st}$. This mechanism explains why  Bose statistics leads 
to the small coefficient of shear viscosity $\eta =G\tau _{st}$.

 When the system is at high temperature, the coherent wave function has a microscopic size.  
 If a long arrow in Fig.2 takes a particle out of the coherent wave 
 function, one cannot regard the particle after  
 displacement as an equivalent of the initial one.  The mechanism below 
$T_{\lambda}$ does not work for the large  displacement extending over two different wave 
functions.  Hence, the relaxation time $\tau _{st}$ changes to an ordinary 
long $\tau _{st}$, which is characteristic of the classical liquid \cite {ber}.

When the system is just above $T_{\lambda}$, the size of the coherent 
wave function is not yet macroscopic, but it develops to an intermediate 
size. In the repulsive system  with high density, the long-distance 
displacements of particles takes much energy,  and therefore particles in 
the low-energy excitation are likely to stay within the same wave function. The excitation energy is 
not so high as that of the  long-distance displacements, but owing to 
Bose statistics, it is not so low as that of the classical liquid. Fast relaxations  
 within these wave functions is the reason for the decrease of $\eta $ in 
 the macroscopic capillary flow.  (In contrast, the ideal Bose system is 
 too simple for such a situation to be realized, in which the macroscopic 
 condensate at $T<T_{\lambda}$ is a necessary condition for the decrease of $\eta $ as in Eq.(26).)

\section{A model of the onset of superfluid flow}
Let us develop a microscopic model of the onset of superfluid flow. 
In deriving the shear viscosity $\eta $, Eq.(4) has the following advantage over Eq.(2). 
In the weak-coupling system (a gas, and a simple liquid like  
liquid helium 4), the particle interaction $U$ normally enhances the 
relaxation of individual particles to local equilibrium positions under a 
given external force, thereby leading to small $\tau _{st}$ and   
$\eta =G\tau _{st}$ \cite {stro}. If we try to formulate this tendency using Eq.(2), that 
is, to derive a decrease of $\eta$ from an increase of $U$ in Eq.(2), there 
must be delicate cancellation of higher-order  
terms in the perturbation expansion of $<J_{xy}(0)J_{xy}(t)>$. 
 On the contrary, if we regard Eq.(4), in which $\eta$ appears in the denominator of $\sigma$,
 as a linear-response relation, we apply the Kubo formula 
 not to $\eta $ but to the {\it reciprocal \/} $1/\eta $.
In the perturbation expansion of $1/\eta$ with respect to $U$, 
{\it an increase of $U$ generally leads to an increase of $1/\eta$ and  
thereby a decrease of $\eta$ \/}.  Hence, one need not expect the 
cancellation. In this respect, when we 
use Eq.(4) as a starting point, the influence of $U$ on $\eta $ is 
naturally built into the formalism from the beginning.

\subsection{The onset of nonclassical behavior}

 For  liquid helium 4, we use the 
 following hamiltonian with the repulsive interaction $U$
\begin{equation}
 H=\sum_{p}\epsilon (p)\Phi_{p}^{\dagger}\Phi_{p}
   +U\sum_{p,p'}\sum_{q}\Phi_{p-q}^{\dagger}\Phi_{p'+q}^{\dagger}\Phi_{p'}\Phi_p , 
   \qquad (U>0),¥¥¥
	\label{¥}
\end{equation}¥
where $\Phi_{p}$ denotes an annihilation operator of a spinless boson. 
Beginning with the repulsive Bose system with no ODLO, we make a perturbation expansion.
 Under the particle interaction $\hat{H_I}(\tau)$, $\chi_{\mu\nu}(q,\omega)$ is derived from  
 \begin{equation}
		<G|T_{\tau}j_{\mu}(x,\tau)j_{\nu}(0,0)|G>  
       =  \frac{\displaystyle{<0|T_{\tau}\hat{j}_{\mu}(x,\tau)\hat{j}_{\nu}(0,0)
 	              exp\left[-\int_{0}^{\beta¥}d\tau \hat{H}_I(\tau)¥\right]|0>¥}}
 	        {\displaystyle{<0|exp\left[-\int_{0}^{\beta¥}d\tau  \hat{H}_I(\tau)¥\right]|0>¥}}¥,
	\label{¥}
\end{equation}¥ 
where $\beta =1/(k_BT), \tau =it$, and $\hat{j}_{\mu}(x,\tau)$ is an interaction representation of Eq.(C2). 
\begin{figure}
\includegraphics [scale=0.59]{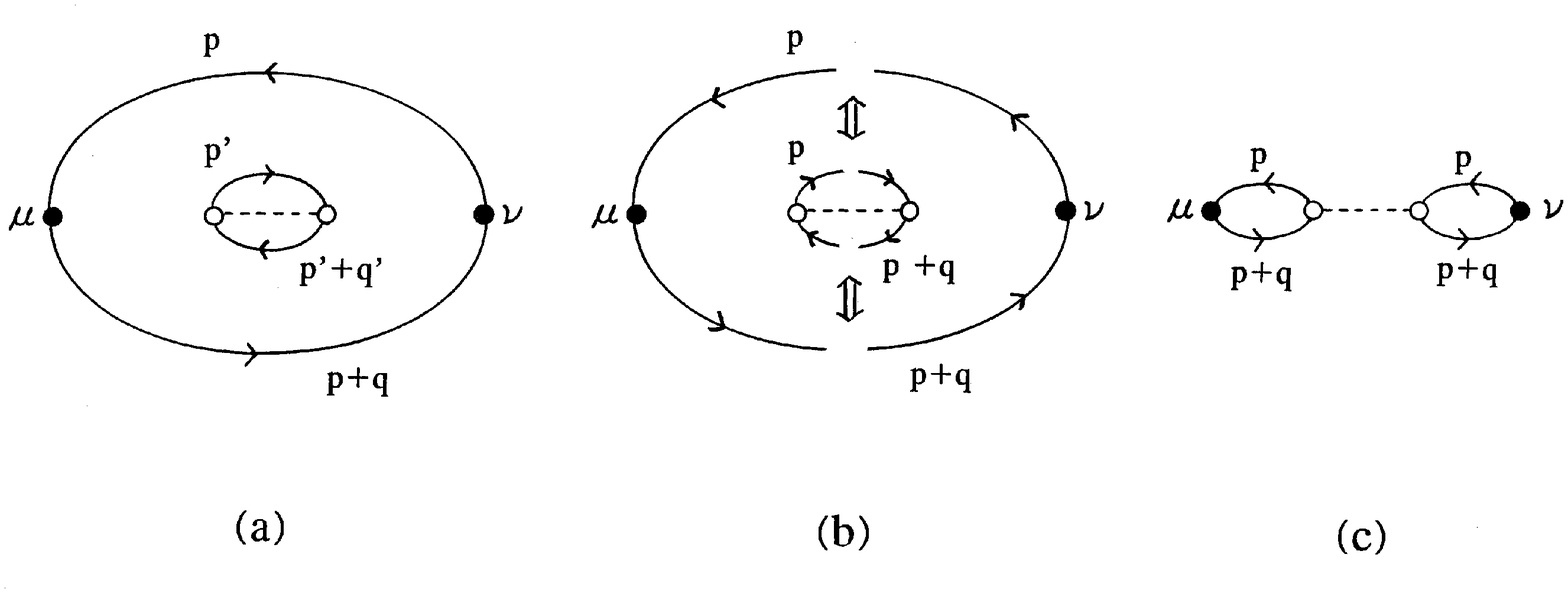}
\caption{\label{fig:epsart} When $p=p'$ and $p+q=p'+q'$ in (a), an 
          exchange of particles in (b) 
         between $j_{\mu}j_{\nu}$ (a large bubble) and a bubble  
           excitation by $\hat{H}_I$ (a small one) yields (c). 
            }
\end{figure}
 The current-current response tensor $\hat{j}_{\mu}(x,\tau)\hat{j}_{\nu}(0,0)$ 
  (a large bubble with $\mu $ and $\nu $ in Fig.4(a)) is in the medium in 
  which particles experience the repulsive interaction:  
  Owing to $exp(-\smallint \hat{H}_I(\tau)d\tau)$ in Eq.(28), 
 scattering of particles frequently occurs in the medium, an example of which is illustrated  by an 
 inner small bubble with a dotted line $U$ in Fig.4(a). As the system approaches $T_{\lambda}$, 
 $\hat{j}_{\mu}(x,\tau)\hat{j}_{\nu}(0,0)$ and $\hat{H}_I(\tau)$ in Eq.(28) 
 get to obey Bose statistics strictly, that is, the particles in the large and small bubbles  
  in Fig.4(a) form a coherent wave function as a whole \cite {com}: 
  When one of the two particles in the large  bubble and in the 
  small bubble have the same momentum 
($p=p'$), and when the other particles in both bubbles have another same momentum 
($p+q=p'+q'$), owing to Bose statistics, a graph made by 
exchanging these particles  must be included in the expansion of Eq.(28). The
cutting and reconnection of these lines between the large and small 
bubbles in Fig.4(b) yields Fig.4(c), in which   
 two bubbles with the same momenta are linked by the repulsive interaction.
 With decreasing temperature, the coherent wave function grows, 
 and such an exchange of particles occurs many times.  
 Furthermore,  with deceasing temperature, particles with the zero momentum get to play a more  
dominant role than others. In Fig.4, this means that taking only processes 
including zero-momentum particles becomes a good approximation:
In $j_{\mu}j_{\nu}$ of Eq.(26), a bubble with $p=0$ corresponds to an excitation from the rest  
particle, and a bubble with $p=-q$ corresponds to a decay into the rest 
one.  Taking only such processes and continuing these exchanges, one obtains  
\begin{equation}
	\hat{\chi}_{\mu\nu}(q,i\omega)=\frac{q_{\mu}q_{\nu}}{4¥}¥
	                 \frac{1}{V¥}\left[	\frac{ F_{\beta}(q,i\omega)}{1-U F_{\beta}(q,i\omega)¥}¥
	                           +\frac{F_{\beta}(q,-i\omega)}{1-U F_{\beta}(q,-i\omega)¥}¥ \right]¥,¥¥
	\label{¥}
\end{equation}¥
where the first and second term in the right-hand side corresponds to the 
$p=0$ and $p=-q$ term in Eq.(26), respectively, and 
\begin{equation}
	 F_{\beta}(q,i\omega)= \frac{(\exp(\beta[\Sigma-\mu])-1)^{-1}-(\exp(\beta[\epsilon (q)+\Sigma-\mu])-1)^{-1}} 
	                         {-i\omega +\epsilon (q)¥¥} ¥.
	\label{¥}
\end{equation}¥

Let us focus on the low-energy excitations. 
For a small $\omega$, the first term in the bracket of the right-hand side 
of Eq.(29) is expanded with respect to $\omega$ as
\begin{eqnarray}
	\frac{ F_{\beta}(q,i\omega)}{1-U F_{\beta}(q,i\omega)¥}¥
	      	     &=&\frac{f(\epsilon (0))-f(\epsilon (q))}{\epsilon (q)-U[f(\epsilon (0))-f(\epsilon (q))]¥}¥\\  \nonumber 
	             &\times &  \left[1+\frac{i\omega}{\epsilon (q)-U[f(\epsilon (0))-f(\epsilon (q))]¥}¥
	                 +\frac{(i\omega)^2}{2(\epsilon (q)-U[f(\epsilon (0))-f(\epsilon (q)] )^2} \cdots¥\right].¥¥
	\label{¥}
\end{eqnarray}¥
Hence, Eq.(29) is expanded as
\begin{equation}
	\hat{\chi}_{\mu\nu}(q,\omega)=\frac{q_{\mu}q_{\nu}}{4¥}¥ \frac{1}{V¥}
	             \frac{ 2F_{\beta}(q,0)}{1-U F_{\beta}(q,0)¥}¥
	    \left[1-\left(\frac{\omega}{\epsilon (q)¥(1-U F_{\beta}(q,0)¥)¥}\right)^2 +\cdots ¥\right]¥.
	\label{¥}
\end{equation}¥

Since the macroscopic flow is of our interest, let us focus on 
$\hat{\chi}_{\mu\nu}(q,\omega)$ at a small $q$.  
 $F_{\beta}(q,0)$ in Eq.(30) is a positive and monotonically decreasing 
function of $q^2$, which approaches zero as $q^2\rightarrow \infty$. 
 An expansion of $F_{\beta}(q,0)$ around $q^2=0$, $F_{\beta}(0,0)-bq^2+\cdots$ has a form such as
\begin{eqnarray}
	 F_{\beta}(q,0) &=&\frac{\beta}{4\sinh ^2 \displaystyle{\left(\frac{|\beta[\mu(T)-\Sigma]}{2¥}\right)}¥¥¥}
	                   \left[1-\frac{\beta}{2¥}\frac{1}{\tanh  
	                   \displaystyle{\left(\frac{|\beta[\mu(T)-\Sigma]|}{2¥}¥\right)}¥¥}
	                        \frac{q^2}{2m¥}¥¥¥¥  +\cdots    \right]¥¥ \\ 
	                &\equiv & a-bq^2 . \nonumber 
	\label{¥}
\end{eqnarray}¥
 For a small $q$, $1-UF_{\beta}(q,0)¥$ in Eq.(32) is approximated as $(1-Ua)+Ubq^2¥$, and we obtain
\begin{equation}
	\hat{\chi}_{\mu\nu}(q,\omega)=\frac{q_{\mu}q_{\nu}}{4¥}¥ \frac{1}{V¥}\frac{2a}{(1-Ua)+Ubq^2¥}¥
	   \left[1-\frac{\omega ^2}{(2m)^{-2}[(1-Ua)q^2+Ubq^4]^2} +\cdots ¥\right]¥.
	\label{¥}
\end{equation}¥
At $T\gg T_{\lambda}$ ($1-Ua>0$), $	\hat{\chi}_{\mu\nu}(q,0)$ is 
approximated as $2aq_{\mu}q_{\nu}/[4V(1-Ua)]$. Hence, the mechanical 
superfluid density $\hat {\rho _s}(T)= \lim _{q\rightarrow 
0}[(q^2/q_{\mu}q_{\nu}) \hat {\chi} _{\mu\nu}(q,0)]$ is zero.  

 With decreasing temperature, the coherent wave function gradually grows. 
 The chemical potential $\mu$, which reflects the size of this wave 
 function \cite {fey2}, gradually approaches $\Sigma$. 
 As $\mu -\Sigma\rightarrow 0$ in Eq.(30), $F_{\beta}(q,i\omega)$ gradually 
increases. Hence, the higher-order terms get to play a dominant role in Eq.(29). 
  As $\mu -\Sigma\rightarrow 0$, $Ua$ ($=UF_{\beta}(0,0)$) in Eq.(34) 
 increases and finally reaches 1, that is,   
\begin{equation}
     U\beta=4\sinh ^2\left(\frac{\beta[\mu (T)-\Sigma(U)]}{2¥}¥\right)¥ .
	\label{¥}
\end{equation}¥
At this point, $\hat{\chi}_{\mu\nu}(q,0)$  has a form of $a/(2VUb)\times q_{\mu}q_{\nu}/q^2$. 
 Hence, the mechanical superfluid density $\hat {\rho _s}(T)$ has a nonzero value $a/(2VUb)$. 
 Using $Ua=1$ and $b$ (Eq.(33)), we obtain   
\begin{equation}
	\hat {\rho _s}(T)=\frac{1}{V¥}\frac{m}{\sinh |\beta [\mu(T)-\Sigma] |¥}¥.
\end{equation}¥
 If we would use the $V\rightarrow\infty $ limit, Eq.(36) is zero unless 
 $\mu -\Sigma =0$. But when we avoid the $V\rightarrow\infty $ limit, it 
 remains finite even when $\mu -\Sigma \ne 0$, and we can estimate $\hat {\rho _s}/\rho$ using 
experimental data  above $T_{\lambda}$ (see Sec.5). Here, we call $T$ satisfying Eq.(35)   
 the onset temperature $T_{on}$ of the nonclassical behavior.

 In the vicinity of $T_{\lambda}$, Eq.(35) is approximated as 
$U\beta=\beta^2[\mu (T)-\Sigma(U)]^2$ for a small $\mu-\Sigma$. This 
 condition has two solutions $\mu (T)=\Sigma(U)\pm \sqrt{Uk_BT}$.  
The repulsive Bose system is generally assumed to undergo BEC 
  as well as a free Bose gas. Hence, with decreasing temperature, 
  $\mu (T)$ of the repulsive Bose system rapidly increases from a 
  negative large value, and reach a positive $\Sigma (U)$ at a finite temperature, 
  during which course the system necessarily passes a 
 state satisfying  $\mu (T)=\Sigma (U)-\sqrt{Uk_BT}¥$. 
 Consequently, $T_{on}$ is always above $T_{\lambda}$, and $\hat {\rho _s}(T)$ has a 
 nonzero value Eq.(36) just above $T_{\lambda}$. This means that 
 $\nu (T)$ in Eq.(25) always differs from $\nu _n$ just above 
 $T_{\lambda}$ as in Fig.1. 
 
 At $T_{\lambda}<T<T_{on}$, Eq.(36) serves as an interpolation formula of  
$\hat {\rho _s}(T)$.  As $T$ approaches $T_{\lambda}$,  $\hat 
{\rho _s}(T)$ in Eq.(36) approaches 
  $(m/V)[\exp (\beta [\mu(T)-\Sigma ])-1]^{-1}$. 
 This means that at $T=T_{\lambda}$, $\hat{\rho _s}(T_{\lambda})$ agrees with the 
conventional thermodynamical superfluid density $\rho _s(T_{\lambda})$, 
and it abruptly reaches a macroscopic number.

 \subsection{Stability of superfluid flow}   
 
To discuss the stability of superfluid flow,  we need a simple form of 
$\hat{\chi}_{\mu\nu}(q,\omega)$, which must be exact at a small $\omega$, and must have a 
 reasonable asymptotic behavior for a large $\omega$.  
  At $T=T_{on}$ ($Ua=1$), we rewrite  Eq.(34) as
 \begin{equation}
		\hat{\chi}_{\mu\nu}(q,\omega)=\frac{q_{\mu}q_{\nu}}{q^2¥}¥\frac{2a^2}{4Vb¥}¥
		 \left[1-\frac{\omega ^2}{B¥}+\cdots 	¥\right]¥,
	\label{¥}
\end{equation}¥
where 
\begin{equation}
	 B=\frac{\epsilon (q)^4}{4(k_BT)^2¥}¥\frac{1}{\tanh ^2 \displaystyle{\left(\frac{|\beta[\mu(T)-\Sigma]|}{2¥}¥\right)}¥}¥.
	\label{¥}
\end{equation}¥
The simplest and practical form of $\hat{\chi}_{\mu\nu}(q,\omega)$ satisfying these conditions is given by
 \begin{equation}
	\hat{\chi}_{\mu\nu}(q,\omega)
	     \simeq \frac{q_{\mu}q_{\nu}}{q^2¥}¥\frac{2a^2}{4Vb¥}¥ 
	                \frac{1}{\left(¥\displaystyle{1+\frac{\omega ^2}{B¥}}¥\right)¥}¥.
	\label{¥}
\end{equation}¥
Using the definition of $\hat{\chi}_{\mu\nu}=(q_{\mu}q_{\nu}/q^2)[\chi^L(q,\omega)-\chi^T(q,\omega)]$,
we find a quantity representing the balance between the longitudinal and transverse responses as follows
\begin{equation}
	\chi^L(q,\omega)-\chi^T(q,\omega)=\frac{1}{V¥}¥
	     \frac{m}{\sinh |\beta [\mu(T)-\Sigma] |¥}¥ \frac{1}{\left(¥\displaystyle{1+\frac{\omega ^2}{B¥}¥}¥\right)}¥.
	\label{¥}
\end{equation}¥
The stability of superfluid flow is measured by Eq.(40). As the Bose 
statistical coherence develops in Eq.(38) ($\mu(T)-\Sigma \rightarrow  
0$), the superfluid flow becomes more stable with respect to higher-frequency external perturbations in 
Eq.(40) ($B \rightarrow \infty $)  \cite {rep}.

The stability of superfluid flow is related to the repulsive interaction 
$U$. In the repulsive system, when a particle is dropped from a superfluid flow 
by external perturbations, such a drop decreases the kinetic energy of the system. 
 However, since that particle does not move similarly to other particles, it raises
the interaction energy between particles, thereby raising its 
total energy. Hence, such a drop from the flow is prevented, and the 
superfluid flow becomes stable. In the famous argument on the stability 
of the superfluid by Landau, the change of one-particle spectrum from $p^2/(2m)$ to $v_sp$ by the 
repulsive interaction is crucial \cite {landau}. 
 We can see the mechanism corresponding to it in Eq.(40) as follows. Even if 
$U=0$ in Eq.(34), $\chi^L(q,\omega)-\chi^T(q,\omega)$ can be written in the 
form of Eq.(40), with $\sinh ^2(|\beta [\mu(T)-\Sigma] |/2)$ replacing 
$\sinh |\beta [\mu(T)-\Sigma] |$. But in this case, $B$ is simply $\epsilon (q)^2$. Hence,  
 with increasing $\omega$, $\chi^L(q,\omega)-\chi^T(q,\omega)$ vanishes 
 far more  rapidly than the case of $B$ in Eq.(38). 
This means that the superfluid flow in the case of $U=0$ is remarkably unstable  
with respect to external perturbations. The repulsive interaction $U$ 
gives $1/\tanh ^2(|\beta[\mu(T)-\Sigma]|/2)$ to $B$ in Eq.(38).
Hence, $\chi^L(q,\omega)-\chi^T(q,\omega)$ does not easily vanish for a large 
$\omega$ in Eq.(40), thereby making the superfluid stable. 
 In this sense, Eqs.(38) and (40) are {\it the many-body theoretical expression\/} of 
the mechanism pointed out by Landau. 
{\it The repulsive interaction plays a significant role both in the emergence 
and in the stabilization of the frictionless flow.\/} 

\subsection{Conductivity of the capillary flow in the vicinity of $T_{\lambda}$ }  
As discussed in Sec.3, what exists behind the suppression of shear 
viscosity near $T_{\lambda}$ is the change of conductivity spectrum $\sigma (\omega)$ near $T_{\lambda}$.  
 In addition to the sharp peak at $\omega =0$, there must be a gradual change in  
$\sigma (\omega)$. Using Eq.(40) in Eq.(20) with the aid of 
\begin{equation}
	\int_{0}^{\infty ¥}¥d\omega\frac{1}{\omega ^2-\omega '^2¥}¥
	\frac{1}{\left(¥\displaystyle{1+\frac{\omega ^2}{B¥}¥}¥\right)}¥
	   = \frac{-\pi}{2\sqrt {B}\left(¥\displaystyle{1+\frac{\omega '^2}{B¥}¥}¥\right)}¥¥,
	\label{¥}
\end{equation}¥
we obtain
\begin{eqnarray}
  \sigma (\omega)&=&\sigma_{1n}(\omega) \\ \nonumber
                  &+&\frac{V^{-1}}{\rho\pi d^2¥}¥\left[\frac{m}{\sinh |\beta [\mu(T)-\Sigma] |¥}¥\delta (\omega )
            -\frac{4}{\pi ^2¥}¥\left(\frac{k_BT}{\epsilon (q_e)^2¥}\right)¥¥
                    \frac{m}{ \displaystyle{\cosh ^2 \left(\frac{|\beta [\mu(T)-\Sigma] |}{2¥}\right)¥¥}¥}¥
           \frac{\pi}{\left(1+\displaystyle{\frac{\omega ^2}{B¥}¥}¥\right)}¥ \right]¥ ,
	\label{¥}
\end{eqnarray}¥
where $q_e$ in $\epsilon (q)$ and $B$ in Eq.(42) is a wave number 
corresponding to the characteristic length of a rotating bucket or a capillary
(from now, we simply denote $\sigma_1(\omega)$ by $\sigma (\omega)$).  
 $\sigma_{1n}(\omega)$ in the right-hand side of Eq.(42) represents the 
 conductivity of a classical liquid ($\sigma (\omega)$ in Fig.3(a)).
The sharp peak and the negative continuous part in the bracket of Eq.(42) 
represent the change of $\sigma (\omega)$ from Fig.3(a) to 3(b).
The total conductivity $\sigma (\omega)$ must satisfy the sum rule 
Eq.(7) regardless of whether it is in the normal or the superfluid phase, 
and therefore we introduced a normalization constant $4/\pi ^2$ in Eq.(42).  
(With the aid of $\int_{0}^{\infty¥}¥d\omega/(1+\omega^2/B)=\pi \sqrt 
{B}/2$, the second term in  the bracket of the right-hand side of Eq.(42) yields a 
term proportional to $1/\sinh |\beta [\mu(T)-\Sigma]| $ in 
$\int_{0}^{\infty¥}¥\sigma (\omega)d\omega$, and $4/\pi ^2$ is determined so that the   
first and second term cancel each other.)

$\sigma _{1n}(\omega )$ in Eq.(42) is given by the real part of Eq.(B5) 
in Appendix.B. Hence, the conductivity spectrum $\sigma _n(\omega )$ of 
the normal-fluid flow has a form such as
\begin{eqnarray}
  \sigma _n(\omega )&=&\frac{1}{\omega d^2¥}Im\left(1-\frac{1}{J_0\left(id[1+i]
                           \displaystyle {\sqrt {\frac{\omega}{2\nu _n¥}¥}}\right)¥¥}\right)¥\\ \nonumber
                    &-&\frac{4}{\pi ^2 nd^2¥}¥\left(\frac{k_BT}{\epsilon (q_e)^2¥}\right)¥¥
                    \frac{V^{-1}}{ \displaystyle{\cosh ^2\left(\frac{|\beta [\mu(T)-\Sigma] |}{2¥}¥\right)¥ }¥}¥
                    \frac{1}{\left(1+\displaystyle{\frac{\omega  ^2}{B¥}¥}¥\right)}¥,	
	\label{¥}
\end{eqnarray}¥
($J_0$ is the zero-th order Bessel function, and $n=\rho /m$).
The temperature dependence comes from second term in the right-hand side 
of Eq.(43).  As $\mu(T)-\Sigma \rightarrow 0$, the second term 
 slightly increases owing to $1/\cosh ^2 |\beta [\mu(T)-\Sigma]/2|$, 
 but  owing to $B\rightarrow \infty $,  $\int_{0}^{\infty¥}¥ \sigma  
 _n(\omega )d\omega$ deceases.  On the other hand, $\sigma _s(\omega )$ of 
the superfluid flow satisfies
\begin{equation}
	\int_{0}^{\infty ¥}¥\sigma _s(\omega )d\omega=\frac{1}{\pi nd^2¥}¥
	     \frac{V^{-1}}{\sinh |\beta [\mu(T)-\Sigma] |¥}¥ ¥.
	\label{¥}
\end{equation}¥
The the sharp peak at $\omega =0$ increases owing to $1/\sinh |\beta [\mu(T)-\Sigma] |$.  
 Equation (43) and (44) describe the change of the conductivity spectrum 
$\sigma (\omega)$ in the vicinity of $T_{\lambda}$ in Fig.3.

\section{Comparison to experiments}
Let us estimate $\hat {\rho¥}_s(T)/\rho$ using Eq.(25) and Fig.1, and compare it to Eq.(36).

(1) Comparing Fig.1 and Eq.(25), we regard $\nu _n$ in the right-hand side of 
Eq.(25) as $\nu (T_{on})$ with $T_{on}=3.7K$,  
and obtain $\hat {\rho¥}_s(T)/\rho$ by
\begin{equation}
 \frac{8}{\pi d^2\omega_e¥}¥\frac{\hat {\rho¥}_s(T)}{ \rho¥}¥ =\nu (T)^{-1}-\nu (T_{on})^{-1}.
	\label{¥}
\end{equation}¥ 
(a) In the experiments  using a capillary, the 
level $h(t)$ of a liquid in the reservoir varies as $h(t)=h(0)e^{-\alpha t}$. In
 Ref.\cite {zin}, this type of experiment was performed in liquid helium 
 3, and $\alpha$ of liquid helium 3 at $1.105K$ was $5\times 
 10^{-4}s^{-1}$, which value must be close to that at $3.7K$. Furthermore, 
 liquid helium 3 and 4 have close values of $\alpha$ under the same 
 experimental condition near $3.7K$ (see Appendix.A.)  Hence, we approximately use 
 $5\times 10^{-4}s^{-1}$ for liquid helium 4 at $3.7K$. 
Using this $\alpha$ in Eq.(5), we estimate the frequency 
width $\omega_e$ of $\Delta P(\omega)$ in $\Delta P(0)e^{-\alpha t}$ to be 
the half width of the sharp peak $1/\sqrt{\omega ^2+\alpha ^2¥}¥$  
at $\omega =0$, obtaining $\omega_e \simeq \sqrt{3}¥\alpha$.  
A rough estimate of $\omega_e$ is $\omega_e \simeq 5\times 10^{-4} rad/s$. 
(b) For the capillary radius $d$, we use a typical value $d\simeq 10^{-2}cm$ 
in Ref.\cite {zin}. (c) In Fig.1, $\nu (T)^{-1}-\nu (T_{on})^{-1}$ reaches $2 \times 10^3 s/cm^2$ 
just above $T_{\lambda}$. Using these $\omega_e$ and $d$ in comparing  Eq.(45) 
to $\nu (T)$ of Fig.1,  we find just above $T_{\lambda}$
\begin{equation}
 \frac{\hat {\rho¥}_s(T)}{\rho¥}¥ = 7\times 10^{-5} .\nonumber   
	\label{¥}
\end{equation}¥  
 Figure.5 shows the temperature dependence of $\hat {\rho¥}_s(T)/\rho$ given by 
 $\nu (T)$ of Fig.1 \cite {dri}.

\begin{figure}
\includegraphics [scale=0.5]{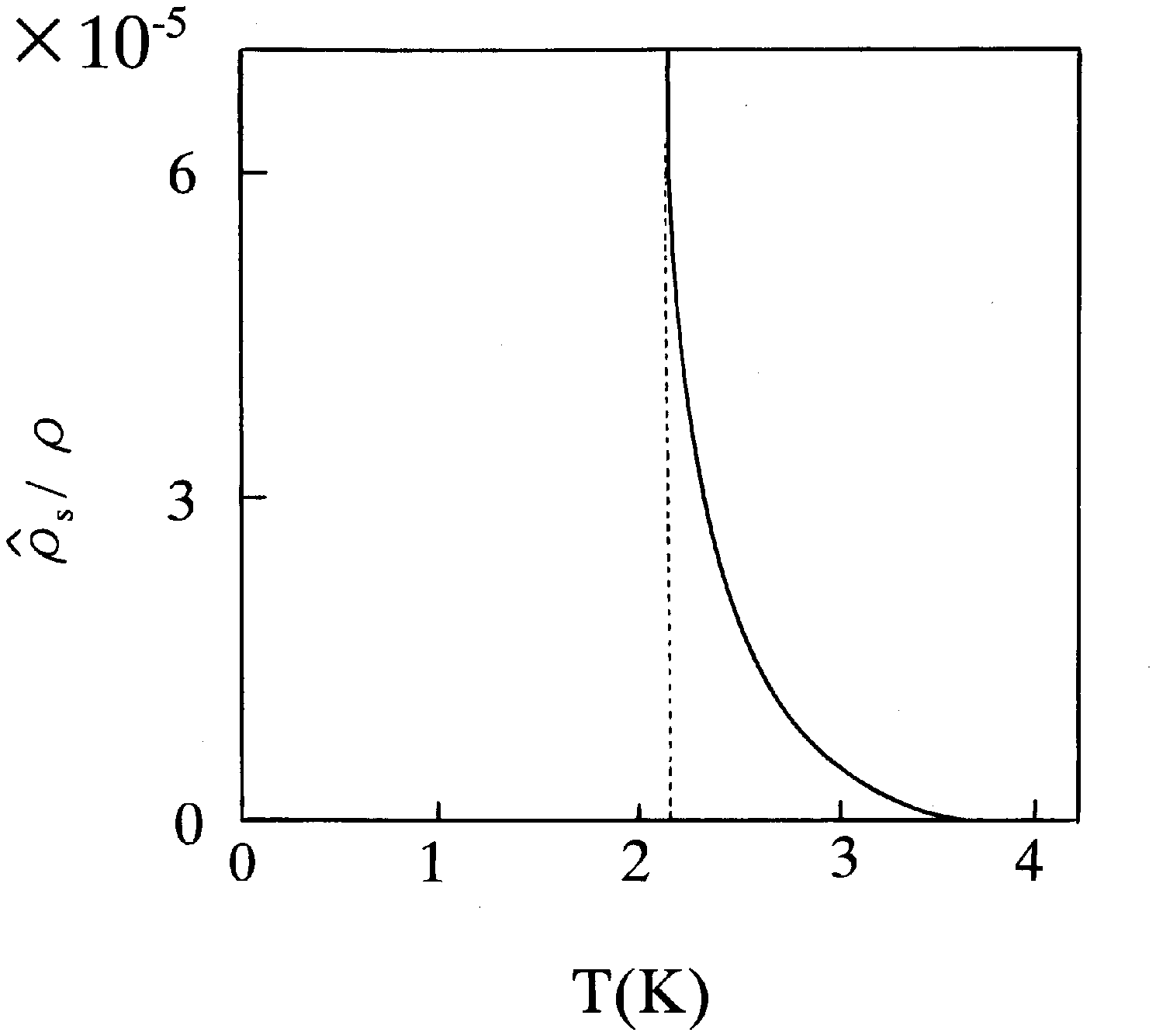}
\caption{\label{fig:epsart} 
 $\hat {\rho¥}_s(T)/\rho$ obtained by Eq.(45) using $\nu (T)$ in 
  Fig.1.  }
\end{figure}

 In the rotation experiment by Hess and Fairbank \cite {hes}, the 
 moment of inertia $I_z$ just above $T_{\lambda}$ is slightly smaller than the  normal 
 phase value $I_z^{cl}$ (see Sec.6.A).  Using these  currently available data,  Ref.\cite {koh} roughly 
 estimates $\hat {\rho¥}_s(T)/\rho$ as $\hat{\rho _s}(T_{\lambda}+0.03K)/\rho\cong 8\times 10^{-5}$, and  
 $\hat{\rho _s}(T_{\lambda}+0.28K)/\rho\cong 3\times 10^{-5}$. Although 
 the precision of these estimates is limited, the agreement of $\hat {\rho¥}_s(T)/\rho$ in Fig.5 
 with these values is good beyond our expectation.

 (2) In Sec.4, we obtained the interpolation formula for $\hat {\rho¥}_s(T)/\rho$, Eq.(36). 
We assume $\mu -\Sigma $ changes with temperature according to the formula
 \begin{equation}
	\mu (T)-\Sigma (U)=-\left(\frac{g_{3/2}(1)}{2\sqrt{\pi}¥}¥\right)^2k_BT_{\lambda}
	          \left[\left(\frac{T}{T_{\lambda}¥}\right)^{3/2}-1\right]^2,
	\label{¥}
\end{equation}¥ 
($g_{a}(x)=\sum_{n}x^n/n^{a}¥$). Here we assume that the particle interaction $U$ and 
the particle density $\rho$ of liquid helium 4 are renormalized to 
$T_{\lambda}=2.17K$, which is an approximation that dates back to London. 
 Equation.(36) with Eq.(46) predicts a temperature  dependence of $\hat {\rho¥}_s(T)/\rho$.  
Although it bears a qualitative resemblance to  
 $\hat {\rho¥}_s(T)/\rho$ in Fig.5, it differs from Fig.5  
 in  that $\hat {\rho¥}_s(T)/\rho$ in Eq.(36) remains very small at $3.7K>T>2.3K$ $(=T_{\lambda}+0.2K)$,
but it abruptly increases at 2.3K, and reaches a 
 macroscopic number at $T_{\lambda}$, thus resembling the shape of the letter $L$. 
   Experimentally, as  temperature decreases from 3.7K to 
$T_{\lambda}$, $\hat {\rho¥}_s(T)/\rho$ gradually increases as in Fig.5. This 
means that Eq.(36) and the approximation behind it is too simple to be 
compared quantitatively to the real system. With decreasing temperature, in addition to the particle  
 with $p=0$, other particles having small but finite momenta get to 
 contribute to the $1/q^2$ divergence of $\hat{\chi}_{\mu\nu}(q,0)$ as 
 well. (In addition to Eq.(30), a new $F_{\beta }(q,0)$ 
 including $p \ne 0$  also satisfies $1-UF_{\beta}(0,0)=0$ in Eq.(29).) 
Hence,  the total $\hat{\rho _s}(T)$ is a sum of each $\hat{\rho _s}(T)$ 
over different momenta \cite {den}.
The participation of $p \ne 0$ particles into $\hat{\rho _s}(T)$ 
is a physically natural phenomenon.  For the  repulsive Bose system, particles are likely to spread 
uniformly in coordinate space due to the repulsive force. This feature 
makes the particles with $p\ne 0 $ behave similarly with other particles, 
especially with the particle having zero momentum.
 {\it If they move at different velocities along the flow direction, the 
 particle density becomes locally high, thus  raising the interaction energy.\/}  This is the 
reason why {\it many particles with $p \ne 0$ participate in the superfluid 
flow even if it is just above $T_{\lambda}$ \/}.

(3) Equations (35) and (46) with $T_{on}=3.7K$ gives us a rough estimate of $U$ as 
$U\sim 3.4\times 10^{-16}erg$. This value  is approximately close to $U_s\cong 
\hbar ^2/(ma^2)=3.4\times 10^{-17}$ $erg$ derived from the scattering length $a=0.7$ 
 $nm$, which is based on the sound velocity $v_s=220$ $m/s$ and the Bogoliubov 
 formula $v_s=(\hbar /m)\sqrt{4\pi na}$.  But this $U$ is somewhat larger than $U_s$.

 \section{Discussion}
 
 \subsection{Superfluidity in the non-dissipative and the dissipative flows}   
  On the onset mechanism of superfluidity, there are physical 
differences between the non-dissipative and the dissipative flows. 

As an example of the non-dissipative flow, we considered the flow in a rotating bucket in Sec.2, 
in which the quantity directly indicating the onset of superfluidity is the moment of inertia
 \begin{equation}
     I_z= I_z^{cl} \left(1-\frac{\hat {\rho _s}(T)}{\rho¥}¥\right)¥ ,¥
\end{equation}¥
where $I_z^{cl}$ is its classical value.

(1) $I_z$ of a superfluid consists only of a linear term of $I_z^{cl}$, 
and $\hat {\rho _s}(T)$ appears  as a correction to its coefficient. 

(2) The change of $\chi ^L-\chi ^T$ directly affects $I_z$ without being enhanced,  
 and therefore just above $T_{\lambda}$, the effect of a small but finite $\hat {\rho 
 _s}(T)$ on $I_z$ is small \cite {koh}. 

For the dissipative flow, we considered the capillary flow, in which the  
quantity directly indicating the onset of superfluidity is the shear 
viscosity. Comparing Eq.(25) to Eq.(47), we note the following features in $\nu (T)$.

(1) $\nu (T)$ of a superfluid has a form of an infinite power series of  
$\nu _n$, and the influence of Bose statistics appears in all 
coefficients of higher-order terms except for the first-order one. This feature does not 
depend on a particular model of a  
liquid, but on the general argument. (On the other hand, the microscopic 
derivation of the value of $\nu _n$ depends on the model of a liquid,  
which is a subject of the liquid theory and beyond the scope of this paper.)

(2) Because of the emergence of the sharp peak due to Eq.(22) in the dispersion integral,
 a small change of $\hat {\rho _s}(T)$ is strongly enhanced to an 
 observable change of $\nu (T)$ in Eq.(25). 

(3) The existence of $1/d^2$ in front of $\hat {\rho _s}(T)/\rho$ in the 
denominator of Eq.(25) indicates that a narrower capillary shows a 
clearer evidence of a frictionless flow.  
 Similarly, the existence of $1/\omega _e$ indicates that the choice of 
 experimental procedure such as the method of applying the pressure to 
 both ends of the capillary affects the temperature dependence of $\nu (T)$ of the capillary flow.
This means that {\it the dissipative phenomena depend on more variables than the 
 non-dissipative ones.\/}

\subsection{Comparison to thermal conductivity}   
There is another type of significant change of conductivity in liquid 
helium 4, an anomalous thermal conductivity at $T<T_{\lambda}$. 
 Under a given temperature gradient $\nabla T$, a heat flow $q$ satisfies 
 $q=-\kappa\nabla T$, where $\kappa$ is  
 the  coefficient of thermal conductivity. In the critical region above 
 $T_{\lambda}$ ($T/T_{\lambda}-1<10^{-3}$), the rapid rise of $\kappa$ is  
 observed, and finally at $T=T_{\lambda}$, $\kappa$ jumps to an at least $10^7$ times 
 higher value \cite {kel}. (This rise is a subject of the fluctuation 
 theory of liquid helium 4 \cite {pha}.) Comparing to $\eta (T)$ which 
 shows a symptom of its fall in $T_{\lambda}<T<3.7K$, 
 $\kappa (T)$ does not show a symptom of its rise in the same temperature region. 
 
 The coefficient of thermal conductivity has a formally 
 similar structure of correlation function to that of the shear 
 viscosity, but they are qualitatively different phenomena.
 While shear viscosity is associated with the transport of momentum (a vector), 
 thermal conductivity is associated with that of energy (a scalar).  For the vector field (velocity 
 field), the direction of vectors has a huge number of possibilities in its spatial 
 distribution.  Hence, among various flow-velocity fields, we  can   
 regard the flow in a rotating bucket as a non-dissipative counterpart to 
 the  capillary flow.  
 On the other hand, for the scalar field (temperature field), the variety 
 of possible spatial distributions is far limited.  A heat flow is 
 always a dissipative phenomenon, and there is no  non-dissipative counterpart.  
 Hence, to the onset of the anomalous thermal conductivity,  
 the mechanism that amplifies the small $\hat {\rho _s}(T)$ to an  
 observable change cannot be applied. 
 This formal difference between shear viscosity  and thermal conductivity 
  is consistent with the experimental difference between $\eta (T)$ and 
  $\kappa (T)$ at $T_{\lambda}<T<3.7K$.

\subsection{Comparison to  Fermi liquids}   
 The fall of the shear viscosity in liquid helium 3 at $T_c$ is  
 a parallel phenomenon to that in liquid helium 4.  The formalism 
in Sec.2 and 3 is applicable to liquid helium 3 as well.  For the behavior above $T_c$, however, 
there is a striking  difference between   liquid helium 3 and 4. 
The phenomenon occurring in fermions in the vicinity of $T_c$ is not a 
gradual growth of the coherent wave function, but a formation of the 
Cooper pairs by two fermions.  
 (This difference evidently appears in the temperature dependence of the 
 specific heat: $C(T)$ of  liquid helium 3 shows only a sharp peak at $T_c$, 
 and does not show a symptom of its rise above $T_c$.)
 Once the Cooper pairs are formed, they are  composite bosons with high 
 density at low temperature,  and immediately jumps to the superfluid state. 
Hence, the shear viscosity of liquid helium 3 shows an abrupt drop at 
$T_c$ without a gradual fall above $T_c$.

 In electron superconductivity, there is an energy gap due to the 
 formation of Cooper pairs. Hence, in its conductivity spectrum $\sigma 
 (\omega )$ at $T<T_{c}$, there is a frequency gap $\omega _g$ near 
 $\omega =0$ \cite {flu}. In superfluid helium 4, due to the  
 acoustic phonon, there is no energy gap, which is consistent with that 
 $\sigma _n(\omega )$ in Eq.(43) is a weakly $\omega$-dependent function.

\appendix

\section{$h(t)=h(0)\exp (-\alpha t)$}
The mass of a liquid passing through a capillary per unit time is 
$Q=2\pi \rho \int_{0}^{d¥}¥rv_z(r)dr$, which is determined by Eq.(3) as  
\begin{equation}
   Q=\rho\frac{\pi d^4}{8\eta¥}¥\frac{\Delta P}{L¥}¥.
\end{equation}¥
The mass of a liquid in the reservoir with a radius $R$, $\rho\pi R^2h(t)$,  
flows out through a capillary at the rate of $Q$. 
The pressure at the upper and lower end of the capillary set up vertically 
 is $1+\rho gh$ and $1$ atm, respectively, thus leading to $\Delta 
P(t)=\rho gh(t)$ in Eq.(A1). The level $h(t)$ of the liquid in the 
reservoir decreases according to 
\begin{equation}
  \rho\pi R^2\frac{dh}{dt¥}¥ =-\frac{\pi d^4}{8\eta¥}¥\frac{\rho ^2 g}{L¥}¥h,
\end{equation}¥
thereby leading to $h(t)=h(0)\exp (-\alpha t)$ with $\alpha =\rho gd^4/(8\eta LR^2)$. 
 For liquid helium 3, $\rho /\eta$ in $\alpha$ is $3.6 \times 10^3$ at 
 $T=1.1K$, and $3.8 \times 10^3$ at $T=3.0K$. 
 For liquid helium 4, $\rho /\eta$ is $4.2 \times 10^3$ at $T=3.0K$.

 \section{Conductivity spectrum $\sigma (\omega)$}
The Stokes equation under the oscillatory pressure gradient $\Delta Pe^{i\omega t}/L$ 
is written in the cylindrical polar coordinate as follows
 \begin{equation}
 \frac{\partial v}{\partial t¥}¥=\nu \left( \frac{\partial }{\partial r^2¥}+ \frac{\partial }{r\partial r¥}\right)v
                              + \frac{\Delta Pe^{i\omega t}}{\rho¥L}¥.
\end{equation}¥
The velocity has the following form
\begin{equation}
 v(r,t)¥=\frac{\Delta Pe^{i\omega t}}{i\omega \rho¥L}+\Delta v(r,t)¥,
\end{equation}¥
under the boundary condition $v(d,t)=0$. In Eq.(B1), $\Delta v(r,t)$ satisfies 
\begin{equation}
 \frac{\partial \Delta v(r,t)}{\partial t¥}¥=\nu \left( \frac{\partial}{\partial r^2¥}
                      + \frac{\partial }{r\partial r¥}\right) \Delta v(r,t)¥,
\end{equation}¥
and therefore $\Delta v(r,t)$ has a solution written in terms of the Bessel function $J_0(i\lambda r)$
with $\lambda =(1+i)\sqrt {\omega /(2\nu)}$. Hence,
\begin{equation}
 v(r,t)=\frac{\Delta Pe^{i\omega t}}{i\omega \rho¥L¥}¥
                       \left(1-\frac{J_0(i\lambda r)}{J_0(i\lambda d)¥}\right)¥.
\end{equation}¥
Using Eq.(B4) at $r=0$, the conductivity spectrum $\sigma (\omega )$, which is defined by 
Eq.(6) as $\rho v(0,t)=\sigma (\omega )d^2 \Delta Pe^{i\omega t}/L$, is given by 
\begin{equation}
 \sigma (\omega )=\frac{1}{i\omega d^2¥}\left(1 
 -\frac{1}{J_0\left(id[1+i]\displaystyle {\sqrt {\frac{\omega}{2\nu¥}¥}}\right)¥¥}\right)¥.
\end{equation}¥
The real part of Eq.(B5)
\begin{equation}
 Re\sigma (\omega )=\frac{1}{\omega d^2¥}
           \frac{\displaystyle 
           {\sum_{n=0}^{\infty¥}¥\frac{1}{(n!)^2¥}¥}\left(\frac{\omega d^2}{4\nu¥}¥\right)^n\cos\frac{n\pi }{2¥}¥¥}
           {\displaystyle {\left[\sum_{n=0}^{\infty¥}¥\frac{1}{(n!)^2¥}¥\left(\frac{\omega d^2}{4\nu¥}¥\right)^n\cos\frac{n\pi }{2¥}\right]^2¥}¥
                   +\displaystyle {\left[\sum_{n=0}^{\infty¥}¥\frac{1}{(n!)^2¥}¥\left(\frac{\omega 
                   d^2}{4\nu¥}¥\right)^n\sin\frac{n\pi }{2¥}\right]^2}}¥.
\end{equation}¥
 gives a curve of $\sigma (\omega)$ in Fig.3(a).  
(Re $\sigma (0)$ in Eq.(B5) agrees with $\rho /(4\eta )$.)  
Furthermore, it determines the conserved quantity $f(d)\propto d^{-2}$ in Eq.(7). 
 ($\nu $ in Eq.(B5) disappears in $\int \sigma (\omega)d\omega$.)

\section{Flow in a rotating bucket}
 The hamiltonian of a liquid in a coordinate system  rotating with the container is 
  $H-\mbox{\boldmath $\Omega$}\cdot \mbox{\boldmath $L$}$, where  
$\mbox{\boldmath $L$}$ is the total angular momentum. 
 The perturbation $H_{ex}=-\mbox{\boldmath $\Omega$}\cdot \mbox{\boldmath $L$}$ 
is cast in the form $-\sum_{i} (\mbox{\boldmath $\Omega$}\times 
\mbox{\boldmath $r$})\cdot \mbox{\boldmath $p$}¥$ , in which  
$\mbox{\boldmath $\Omega$}\times \mbox{\boldmath $r$} \equiv \mbox{\boldmath $v$}_d( \mbox{\boldmath $r$})$ 
serves as an external field. The rotation is equivalent to the 
application of the external field. We define the mass-flow density 
$\mbox{\boldmath $j$}(\mbox{\boldmath $r$})$ in the rotation, and express  the 
perturbation $H_{ex}$ as
\begin{equation}
   -\mbox{\boldmath $\Omega$}\cdot \mbox{\boldmath $L$}
         =-\int  \mbox{\boldmath $v$}_d( \mbox{\boldmath $r$})\cdot 
   \mbox{\boldmath $j$}(\mbox{\boldmath $r$}) d^3x.¥
\end{equation}¥
Because of $ div\mbox{\boldmath $v$}_d( \mbox{\boldmath $r$})=0 $, 
 $\mbox{\boldmath $v$}_d( \mbox{\boldmath $r$})$ in Eq.(C1)
 acts as a transverse-vector probe to the excitation of bosons. This fact 
 allows us a formal analogy that the response of the system to 
 $\mbox{\boldmath $v$}_d( \mbox{\boldmath $r$})$ is analogous to the 
 response of the charged Bose system to the vector potential 
 $\mbox{\boldmath $A$}(\mbox{\boldmath $r$})$ in the Coulomb gauge.  
 Hence,  in momentum space, $ \mbox{\boldmath $j$}(\mbox{\boldmath $r$})$ in Eq.(C1) has the 
 following microscopic form similar to that in the charged Bose system
 \begin{equation}
	j_{\mu}(q,\tau)=\sum_{p,n} 
	\left(p+\frac{q}{2¥}\right)_{\mu}\Phi_p^{\dagger}\Phi_{p+q}e^{-i\omega _n\tau}¥¥¥,
	\label{¥}
\end{equation}¥
($\hbar =1$ and  $\tau =it$).

\section{Maxwell's relation}
Consider the shear deformation of a solid and of a liquid. In a solid,  
shear stress $F_{xy}$ is proportional to a shear angle $\phi$ as 
$F_{xy}=G\phi$, where $G$ is the modulus of rigidity. The value of $G$ is 
determined by dynamical processes in which vacancies in a solid move to neighboring 
positions over the energy barriers. As  $\phi$ increases, $F_{xy}$ 
increases as follows,
\begin{equation}
 \frac{dF_{xy}}{dt}=G\frac{d\phi}{dt¥}¥¥.
	\label{¥}
\end{equation}¥

\begin{figure}
\includegraphics [scale=0.6]{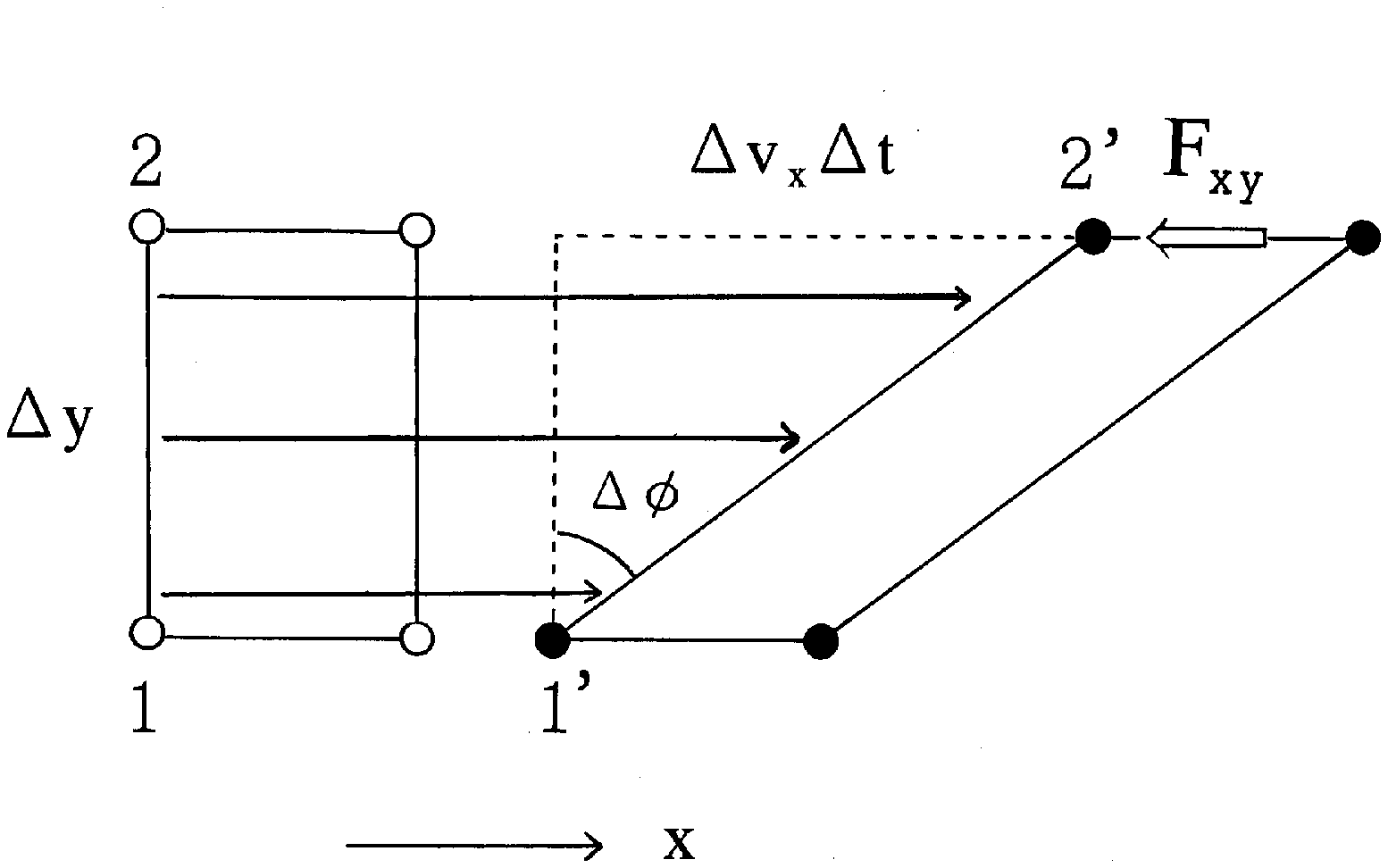}
\caption{\label{fig:epsart} In a liquid  flowing along the $x$-direction, owing to the
 velocity gradient along the $y$-direction, a small rectangular part of a liquid 
 is deformed to a parallelogram. }
\end{figure}
In a liquid, the situation is different as follows. Figure 6 represents a 
small portion of the flow in Fig.2, in which two 
particles 1 and 2, each of which starts at $(x,y)$ and $(x,y+\Delta y)$  
simultaneously,  moves along the $x$-direction. In a liquid, 
the relative position of particles is not rigid,  
but with the flow motion it changes so as to reduce the shear stress 
$F_{xy}$ between adjacent layers.  Presumably, the 
larger shear stress induces the faster structural relaxation, that is, the rate of such relaxations is 
proportional to the magnitude of $F_{xy}$. Hence, instead of Eq.(D1),  one obtains 
\begin{equation}
 \frac{dF_{xy}}{dt}=G\frac{d\phi}{dt¥}-\frac{F_{xy}}{\tau _{st}¥}¥¥¥,
	\label{¥}
\end{equation}¥
where $\tau _{st}$ is a relaxation time. 
In the stationary flow after relaxation, $F_{xy}$ remains constant, and one obtains
\begin{equation}
 G\frac{d\phi}{dt¥}=\frac{F_{xy}}{\tau _{st}¥}¥¥¥.
	\label{¥}
\end{equation}¥
  Assume that there is velocity gradient $v_x(y)$ along $y$ direction in Fig.6. After $\Delta t$ has 
passed, they (1' and 2') are at a distance of $\Delta v_x\Delta t$ along the 
$x$-direction. The shear angle increases from zero to $\Delta\phi$, which satisfies 
$\Delta v_x\Delta t=\Delta y\Delta\phi$ as depicted in Fig.6. Hence, we obtain
\begin{equation}
 \frac{\partial v_x}{\partial y¥}=\frac{d\phi}{dt¥}¥.
	\label{¥}
\end{equation}¥
Substituting Eq.(D4) into Eq.(D3), and comparing it with Eq.(1), one obtains 
$\eta=G\tau _{st}$ (Maxwell's relation).


\begin{thebibliography}{99}
\bibitem{cri} The increase of $\tau _{th}$ by the critical slowing down is negligible at $T_{\lambda}<T<3.7K$.
\bibitem{even} The fluctuation-dissipation theorem is not applicable to the non-dissipative response 
             such as the onset of superfluidity.
\bibitem{han} In the short time scale, the mechanism of the 
           shear viscosity in a liquid is similar to that in a solid, and is therefore  
          a highly inhomogeneous process at the molecular level: The modulus of rigidity $G$ 
          is determined by the motion of vacancy as in a solid.
\bibitem{cur} As a response to $\mbox{\boldmath $P$}/L$, 
           $\mbox{\boldmath $j$}(\mbox{\boldmath $r$})$  at another 
           point is possible, but it results in the same form of $\eta$.
\bibitem{vel} $\mbox{\boldmath $v$}(\omega)$ in the right-hand side of Eq.(10) is a 
         fictitious velocity that appears when the system would obey Eq.(9) for 
         the non-interacting system, whereas $ \mbox{\boldmath $v$}(r=0)$ 
         in $\mbox{\boldmath $j$}(\omega)$ of the left-hand side 
         is a real velocity in the interacting system.
\bibitem{def} We usually use the density $\rho$, a longitudinal susceptibility 
            satisfying $\rho=mn=\chi^L(q,0)$, for the transverse response of the system as in the 
           conductivity $\sigma_{1n}=\rho /(4\eta )$.
           This is explained by the replaceability of $\chi^T$ by $\chi^L$ in Eq.(19).
\bibitem{she} In a classical liquid, $\eta $ is inversely proportional to the rate of
           processes in which  vacancies in a liquid propagate from one point to 
          another over the energy barriers. With decreasing temperature, this rate 
          decreases. The liquid therefore slowly respond to perturbations, and its $\eta $ increases.
\bibitem{stro} In the strong-coupling system, however, the strong interaction 
                    disturbs the relaxation, thereby leading to a large $\eta$.
\bibitem{com} It is possible that  a more complex diagram than a bubble exchanges 
             particles with the tensor $j_{\mu}j_{\nu}$.  While it is 
             difficult to estimate an infinite sum of these diagrams, it adds only a small correction.
\bibitem{dri} The $\hat {\rho¥}_s(T)$ at $T_{\lambda}<T<3.7K$ 
          in Fig.5 is a completely different concept from the fluctuation-driven superfluid density,
           which was proposed in Ref.\cite {pha} at $|T/T_{\lambda}-1|<10^{-2}$.
\bibitem{den} The density of states for this sum is not proportional
           to $4\pi p^2$ but a constant, because it is owing to the one-dimensional momenta along the flow.
\bibitem{flu} In electron superconductors, in addition to the normal  
             conductivity  $\sigma _n$, an extra $\sigma '$ 
             ($\sigma '/\sigma _n \simeq 10^{-5}$) is observed at $|T/T_{c}-1|<10^{-2}$, which 
              is considered to be driven by fluctuations.
\end{thebibliography}

\end{document}